\documentclass[onecolumn,pra,aps,amsmath,amssymb,superscriptaddress]{revtex4}

\usepackage{lineno}
\usepackage{mathptmx}
\usepackage{subfigure}
\usepackage{psfrag,graphicx}
\usepackage{dcolumn}
\usepackage{amsmath,amssymb}
\usepackage{bm}
\usepackage{color}
\usepackage{overpic}
\usepackage{latexsym}
\usepackage{epstopdf}
\usepackage{color}
\usepackage[english]{babel}
\usepackage{latexsym}
\usepackage{stmaryrd}
\usepackage{amsfonts}
\usepackage{psfrag,graphicx} 
\usepackage{epsf} 
\usepackage{subfigure} 
\usepackage{amsmath} 
\usepackage{amssymb} 
\usepackage{amsfonts}
\usepackage{bm}
\usepackage{natbib}

\usepackage{epstopdf}
\usepackage{url}
\usepackage{appendix}
\DeclareMathOperator{\Tr}{Tr}
\newcommand\bigzero{\makebox(0,0){\text{\huge0}}}
\definecolor{mygrey}{gray}{0.35}
\definecolor{myblue}{rgb}{0.2,0.2,0.8}
\definecolor{myzard}{cmyk}{0,0,0.05,0}
\definecolor{mywhite}{rgb}{1,1,1}
\definecolor{myred}{rgb}{1,0.,0.3}

\usepackage[colorlinks=true,citecolor=myblue,linkcolor=myred]{hyperref}
\def\be{\begin{equation}}
\def\ee{\end{equation}}
\def\ba{\begin{align}}
\def\enda{\end{align}}
\def\bi{\begin{itemize}}
\def\ei{\end{itemize}}

 \def\ee{\mathord{\rm e}}

 \def\ee{\mathord{\rm e}}

\renewcommand{\ee}{{\rm e}}

\def\beq{\begin{equation}}
\def\beq{\begin{equation}}
\def\eeq{\end{equation}}

 \newcommand{\ket}[1]{|#1\rangle}

\begin{document}


\title[Short Title]{Off-Diagonal Observable Elements from Random Matrix Theory: Distributions, Fluctuations, and Eigenstate Thermalization}
\author{Charlie Nation}%
 \email{C.Nation@sussex.ac.uk}
\affiliation{%
Department of Physics and Astronomy, University of Sussex, Brighton BN1 9QH, United Kingdom.
}%
\author{Diego Porras}%
\email{D.Porras@sussex.ac.uk}
\affiliation{%
 Department of Physics and Astronomy, University of Sussex, Brighton BN1 9QH, United Kingdom.
} 
\date{\today}

\begin{abstract}
We derive the Eigenstate Thermalization Hypothesis (ETH) from a random matrix Hamiltonian by extending the model introduced by J. M. Deutsch [Phys. Rev. A 43, 2046 (1991)]. 
We approximate the coupling between a subsystem and a many-body environment by means of a random Gaussian matrix. We show that a common assumption in the analysis of quantum chaotic systems, namely the treatment of eigenstates as independent random vectors, leads to inconsistent results. However, a consistent approach to the ETH can be developed by introducing an interaction between random wave-functions that arises as a result of the orthonormality condition. 
This approach leads to a consistent form for off-diagonal matrix elements of observables. From there we obtain the scaling of time-averaged fluctuations of generic observables with system size for which we calculate an analytic form in terms of the Inverse Participation Ratio. 
The analytic results are compared to exact diagonalizations of a quantum spin chain
for different physical observables in multiple parameter regimes.
\end{abstract}

\maketitle
\setcounter{tocdepth}{2}
\begingroup
\hypersetup{linkcolor=black}
\tableofcontents
\endgroup

\section{Introduction}

The emergence of statistical physics from unitary quantum dynamics has been debated since the early days of quantum theory \cite{Neumann10}.  
It is by now widely accepted that generic non-integrable quantum systems undergo a process known as 
quantum thermalization, which implies that an initially out-of-equilibrium state of an isolated quantum system will 
approach thermal equilibrium after some typical relaxation time.
The underlying mechanism behind quantum thermalization is still a subject of debate 
\cite{Reimann2008,Reimann2010,Short2011,Short2012,Linden2009,Ikeda2011,Rigol2012,DAlessio2016,Luitz2016,Gogolin2016,Borgonovi2016,Shiraishi2017,Mori2017}.
One of the most successful approaches to this long-standing problem is the Eigenstate Thermalization 
Hypothesis (ETH)\cite{Srednicki1994, Deutsch1991, Rigol2008}. 
According to this conjecture, the many-body eigenstates of a non-integrable Hamiltonian yield the same 
expectation values of local observables as those calculated with a microcanonical ensemble. 
%
%
Below we will give a more detailed presentation of this conjecture, which can be formulated as an ansatz for the matrix elements of observables in the eigenbasis of a many-body Hamiltonian. 
To visualize qualitatively the physics behind the ETH, we can consider a quantum lattice system with interactions coupling different sites.
If we express a many-body eigenstate in a local basis, we expect that interactions lead to a highly 
entangled state distributed over the lattice\cite{Pechukas1983, Pechukas1984}. 
The ETH assumes that the resulting linear superposition has similar properties to a microcanonical 
ensemble. 
Note that this mechanism for thermalization is purely quantum mechanical since the existence of quantum correlations and entanglement are essential ingredients.

The validity of the ETH has been confirmed for a wide range of non-integrable systems by means of exact 
diagonalizations 
\cite{Santos2010,Santos2010a,Steinigeweg2013,Beugeling2014,Beugeling2015,Hunter-Jones2017, Mondaini2016}. 
Still, there are some aspects of quantum thermalization and the ETH that are not completely clear. 
The conjecture can be qualitatively justified by using the theoretical framework of quantum chaos, however, it has not yet been fully derived mathematically from first principles.
A possible direction to address the validity of the ETH is to try to derive it from a more basic or fundamental assumption or set of assumptions.
In particular, we know that many-body eigenstates of large systems can be often described by Random 
Matrix Theory (RMT). 
The original work by J. Deutsch \cite{Deutsch1991} actually used a Random Matrix Hamiltonian as a 
toy model to show the emergence of quantum thermalization in isolated quantum systems. 
In Deutsch's approach a non-ergodic system is perturbed by a Gaussian random matrix, which results in 
an approximate description of many-body eigenstates by random wave-functions 
with uncorrelated random coefficients. This theoretical framework does not by itself prove the occurrence of thermalization in particular many-body systems, however, it proves certain aspects of the process as long as reasonable assumptions on the underlying system are fulfilled.

The quantum thermalization process has two fundamental aspects. Firstly, it involves the equivalence 
between time-averages of expectation values of observables and microcanonical averages. 
Secondly, it also involves the equilibration of an initially excited state into a thermal state, that 
is, we expect that time-fluctuations around thermal averages will be small. Furthermore, those 
fluctuations should decrease with system size such that statistical mechanics is recovered in the 
thermodynamic limit. 
Equilibration is governed by the off-diagonal matrix elements of an observable in the basis of 
eigenstates of the Hamiltonian. 
The ETH as formulated by Srednicki \cite{Srednicki1994} includes a condition for off-diagonal matrix elements, which ensures equilibration. 
Furthermore, it has been proved that the random wave-function model can be used to qualitatively reproduce the ETH result for time-fluctuations \cite{Reimann2015}. 
However, random wave-function models usually work under the assumption of statistically independent random coefficients. This condition limits the validity of this approach, as we show in the next section. 

A deeper understanding of time-fluctuations in the quantum thermalization process is actually crucial 
to describe current experiments with microscopic systems. 
Physical realizations of isolated quantum systems, where the emergence of statistical physics can be investigated, have been made possible only recently due to advances in quantum simulators with atomic and solid-state systems \cite{Cirac2012,Schneider2012,Nori2014}.
These include ultracold atoms \cite{Gring2012,Schreiber2015,Kaufman2016}, trapped ions \cite{Smith2016,Clos2016a}, and superconducting qubits \cite{Neill2016}. 
Identifying quantum thermalization would ideally involve a comparison between observed time-averages 
and microcanonical averages, however  computing the latter is a challenge in complex many-body systems. 
An alternative path to test theoretical ideas such as the ETH is to check predictions made on the 
time-fluctuations of observables, such as the scaling with system size or interaction strength. 
For that aim, a deeper understanding of the physics and assumptions underlying the ETH would be required, to obtain quantitative predictions that can be used to identify ergodic phases in experiments.

In this work we present a derivation of the ETH in a random matrix model that yields an 
approximate description of a quantum non-integrable system under some reasonable assumptions. 
We build on the theoretical model introduced by J. Deutsch \cite{Deutsch1991}, and extend it to the calculation of off-diagonal matrix elements of observables. 
We show that correlations induced by orthonormality between random wave-functions must be taken into account to obtain a consistent derivation of the ETH from Random Matrix Theory.
Our work cannot be considered as a proof of the validity of the ETH, however, it shows that the 
conjecture can be fully obtained from a description in terms of random wave-functions.
Our theory can be used to quantify time-fluctuations after a quantum quench, and to predict the scaling of fluctuations with system size, thus yielding predictions that can be compared with experimental results and used to identify ergodic regimes in quantum many-body systems.  

This article is structured as follows. In Section II we introduce the ETH ansatz and discuss the limitations of a model of independent random wave-functions to describe the behaviour of off-diagonal matrix elements of observables. In Section III we introduce Deutsch's random matrix model consisting of a diagonal Hamiltonian perturbed by a Gaussian random matrix. 
We extend the original model to account for interactions between random wave-functions arising from the orthonormality condition. In Section IV we calculate the correlation functions between random wave-functions. 
In Section V we use those correlation functions to calculate the off-diagonal matrix elements of an operator, and show that they take the same form predicted by the ETH. 
In Sections VI and VII we present a numerical confirmation of our analytical results. 
Finally, in Section VIII we show that our model provides us with a good description of the 
time-fluctuations in a non-integrable quantum spin chain. We finish with our Conclusions in Section IX, where we discuss the range of applicability of our results and their implications.

\section{Eigenstate Thermalization Hypothesis and the Limitation of the Independent Random Wavefunction Ansatz}

In this section we introduce the ETH and the random wave-function ansatz. We will show that a description of many-body wave-functions based on independent random variables does not lead to a consistent description of off-diagonal matrix elements of typical observables.
 
To focus our discussion, consider a system described by a non-integrable Hamiltonian, $H$, with 
eigenvectors and eigenenergies $|\psi_\mu\rangle$ and $E_\mu$, respectively, such that 
$H | \psi_\mu \rangle = E_\mu | \psi_\mu \rangle$. 
The system is initially in the state $|\Psi(0) \rangle = \sum_\mu a_\mu | \psi_\mu \rangle $ with mean 
energy $\bar{E} := \langle \Psi(0) | H | \Psi(0) \rangle$. 
%
%
The equilibration of a closed quantum system into a thermal state implies that (assuming non-degenerate energy levels),
\begin{equation}
\lim_{T\to \infty}\frac{1}{T} \int_0^T \langle O(t) \rangle dt 
= \sum_{\mu} |a_\mu|^2 O_{\mu \mu} \approx \langle O \rangle_{\rm micro},
\label{thermalization}
\end{equation}
where we have used the definition $O_{\mu\mu} := \langle\psi_\mu|O|\psi_\mu\rangle$.
Eq. (\ref{thermalization}) expresses an equivalence between the time-average of $\langle O(t) \rangle$ and the microcanonical average of $O$ taken over an energy shell of eigenstates with energies $E_\mu$ close to $\bar{E}$.
The ETH for diagonal elements of observables consists of the assumption that $O_{\mu \mu}$ is a smooth function of the energy $E_\mu$, ${\cal O}(E_\mu)$,
\begin{equation}
O_{\mu \mu}  \ \underset{\mathrm{ETH}}{=}  \ {\cal O}(E_\mu).
\label{eth.diag}
\end{equation}
Assuming that probabilities $|a_\mu|^2$ take non-vanishing values close to $\bar{E}$, the ETH ensures that the second term in Eq. (\ref{thermalization}) is equivalent to a microcanonical average.
%

To understand the relation between the ETH and a random wave-function ansatz, let us assume that the observable $O$ is a local operator in a quantum lattice model defined on a subsystem $S$. 
The rest of the lattice forms a bath, $B$, and we write the total Hamiltonian like 
$H = H_S + H_B + H_{SB}$, where $H_{SB}$ is the interaction term. 
Now we define $H_{0} = H_S + H_B$, and the non-interacting energy eigenbasis, 
$H_0 |\phi_\alpha\rangle = E_\alpha |\phi_\alpha \rangle$. 
To simplify the notation in what follows we will assume that variables with indices $\mu$, $\nu$ refer to eigenenergies or eigenstates of the interacting Hamiltonian, whereas indices $\alpha$, $\beta$ refer to $H_0$.

The random wave-function ansatz consists of the assumption that
\begin{equation}
|\psi_\mu \rangle = \sum_{\alpha} c_\mu(\alpha) | \phi_\alpha \rangle ,
\end{equation} 
with $c_\mu(\alpha)$ independent normalized random variables with average
\begin{equation}
\langle c_\mu(\alpha) c_{\mu'}(\alpha')\rangle_V = \delta_{\mu,\mu'} \delta_{\alpha,\alpha'} \Lambda(\mu, \alpha) ,
\end{equation}
where $\Lambda(\mu, \alpha)$ is a function of $(E_\mu - E_\alpha)$, normalized such that $\sum_{\alpha} \Lambda(\mu, \alpha) = \sum_{\mu} \Lambda(\mu, \alpha) = 1$. The average $\langle \cdots \rangle_V$ is taken over realizations of the random wave-function (this will be more clearly defined in the next section). We assume that the function $\Lambda(\mu,\alpha)$ is smooth, has a maximum when $E_\mu = E_\alpha$, and vanishes when $E_\mu - E_\alpha \gg \Gamma$, with $\Gamma$ being a typical energy width. A perturbative calculation, in which $H_{\rm SB}$ was approximated by a random matrix, carried out by Deutsch\cite{Deutscha} leads to a random wave-function model with a Lorentzian, 
\begin{equation}
\Lambda(\mu, \alpha) = \frac{\Gamma \omega_0/\pi}{(E_\mu-E_\alpha)^2 + \Gamma^2},
\end{equation}
where $\omega_0$ is the average spacing between energy levels and we assume for now that both $\Gamma$ and $\omega_0$ are independent of $\alpha$, $\mu$. 
Outside a perturbative regime, however, numerical calculations on non-integrable models have shown that wave-functions have a Gaussian shape \cite{Mondaini2017, Santos2012, Njema1996, Flambaum1998, Atlas2017}.

Diagonal matrix elements in the interacting basis can be approximated under the assumption of self-averaging,
\begin{equation}
O_{\mu \mu} = \sum_{\alpha\beta} c_\mu(\alpha) c_\mu(\beta) O_{\alpha \beta} 
\approx \sum_{\alpha} \Lambda(\mu, \alpha) O_{\alpha \alpha} ,
\label{diagonal}
\end{equation}
where $O_{\alpha\beta} := \langle \phi_\alpha|O|\phi_\beta \rangle $. Eq. (\ref{diagonal}) 
implies that the coupling induced by $H_{\rm SB}$ leads to the smoothing of the distribution of 
diagonal matrix elements in the interacting basis and provides us with a justification for the ETH  for diagonal elements of observables \eqref{eth.diag} within the random wave-function model \cite{Deutsch1991,Reimann2010}, since we can make the identification,
\begin{equation}
{\cal O} (E_\mu) = \sum_{\alpha} \Lambda(\mu, \alpha) O_{\alpha \alpha} ,
\end{equation}
which yields a smooth function as long as the sum runs over a sufficiently large number of states.

We also expect that in the thermodynamic limit the average $\langle O(t) \rangle$ does not deviate too much from its mean-value (equilibration aspect of quantum thermalization). 
The averaged time-fluctuations over an infinite integration time are given by \cite{Srednicki1999},
\begin{equation}
\delta_O^2(\infty) = \sum_{\substack{\mu, \nu \\ \mu \neq \nu}} |O_{\mu \nu}|^2 |a_\mu|^2 |a_\nu|^2,
\end{equation}
under the assumption of non-degenerate energy gaps. Based on quantum chaos theory, Srednicki \cite{Srednicki1994} introduced the ETH ansatz for the  off-diagonal matrix elements $O_{\mu \nu}$,
\begin{equation}
\label{eq:offdiagETH}
O_{\mu \nu} |_{\mu \neq \nu} \underset{\mathrm{ETH}}{=} \frac{1}{\sqrt{D(E)}} f_O(E,\omega) R_{\mu \nu} .
\end{equation}
In this expression $D(E)$ is the density of states (the original expression by Srednicki included an equivalent normalization by using the microcanonical entropy instead), $E = (E_\mu + E_\nu)/2$ and $\omega = E_\nu - E_\mu$. 
$R_{\mu \nu}$ is a set of random variables with zero average and unit variance. 
$f_O(E,\omega)$ is a continuous function of $E$ and $\omega$, which we expect to be centered around $\omega = 0$, and take negligible values if the difference between energies $\omega$ is larger than a typical energy width.

A natural question is whether a random wave-function model can be used to justify the ETH ansatz for off-diagonal matrix elements as well. 
As the off diagonal elements of a typical observable average to zero, it is convenient instead to analyze the squared modulus. To simplify the discussion we restrict our evaluation for now to those observables that are diagonal in the basis of $H_0$,
\begin{equation}{\label{eq:off_diags1}}
|O_{\mu\nu}|^2_{\mu \neq \nu} = \sum_{\alpha\beta}c_\mu(\alpha)c_\nu(\alpha)c_\mu(\beta)c_\nu(\beta)O_{\alpha\alpha}O_{\beta\beta}.
\end{equation}
A common assumption that is made here \cite{Deutsch1991, Reimann2015} is to treat the coefficients $c_\mu(\alpha)$ as uncorrelated random numbers, the only surviving terms of this sum will then be
\begin{equation}{\label{eq:URN_1}}
|O_{\mu\nu}|^2_{\mu \neq \nu} = \sum_{\alpha}|c_\mu(\alpha)|^2|c_\nu(\alpha)|^2O_{\alpha\alpha}^2 \approx \sum_{\alpha} \Lambda(\mu, \alpha) \Lambda(\nu, \alpha) O_{\alpha \alpha}^2,
\end{equation}
which actually agrees with the ETH ansatz. However this expression cannot provide us with a consistent description on off-diagonal matrix elements. To show this, consider the equality
\begin{equation}{\label{eq:sum_nu1}}
\sum_{\nu}|O_{\mu\nu}|^2 = \langle \psi_\mu|O^2|\psi_\mu\rangle.
\end{equation}
Now, analyzing the diagonal and off-diagonal terms of Eq. (\ref{eq:sum_nu1}) separately, \emph{i.e} $\sum_{\nu}|O_{\mu\nu}|^2 = |O_{\mu\mu}|^2 + \sum_{\nu \neq \mu}|O_{\mu\nu}|^2$, for the off-diagonal terms we have, given Eq. (\ref{eq:URN_1}),
\begin{equation}\label{eq:sum_nu_RW}
\sum_{\nu} |O_{\mu\nu}|^2_{\mu \neq \nu} \approx \sum_{\alpha} \Lambda(\mu, \alpha) O_{\alpha \alpha}^2 \approx \overline{O_{\alpha\alpha}^2},
\end{equation}
where we have defined a microcanonical average around $E_\mu$, $\overline{O_{\alpha\alpha}^2}$. 
Now, the sum of off-diagonal elements may also be obtained without use of the random wave-function ansatz as
\begin{equation}\label{eq:O2_munu}
\begin{split}
\sum_{\nu}|O_{\mu\nu}|^2_{\mu \neq \nu} &= (O^2)_{\mu\mu} - |O_{\mu\mu}|^2 \\ &
\approx \sum_{\alpha} \Lambda(\mu, \alpha)O_{\alpha\alpha}^2 - |\sum_{\alpha}\Lambda(\mu, \alpha)O_{\alpha\alpha}|^2 \\&
\approx \overline{O_{\alpha\alpha}^2} - \overline{O_{\alpha\alpha}}^2,
\end{split}
\end{equation}
where we have only assumed a self-averaging condition. Thus, comparing Eq. (\ref{eq:O2_munu}) and Eq. (\ref{eq:sum_nu_RW}) we can observe that we obtain an inconsistency. We are thus lead to conclude that there are indeed correlations between the coefficients, and that Eq. ({\ref{eq:URN_1}}) is na\"{i}ve. If instead we write
\begin{equation}{\label{eq:sum_ave2}}
\sum_{\nu \neq \mu} |O_{\mu\nu}|^2 = \sum_{\nu \neq \mu}\sum_{\alpha}|c_\mu(\alpha)|^2|c_\nu(\alpha)|^2O_{\alpha\alpha}^2  + \sum_{\nu \neq \mu}\sum_{\substack{\alpha\beta\\\alpha \neq \beta}} c_\mu(\alpha)c_\nu(\alpha)c_\mu(\beta)c_\nu(\beta)O_{\alpha\alpha}O_{\beta\beta},
\end{equation}
we can then apply the self-averaging assumption once more, \emph{i.e} the replacement $\sum_{\nu \neq \mu}\sum_{\substack{\alpha\beta\\\alpha \neq \beta}} c_\mu(\alpha)c_\nu(\alpha)c_\mu(\beta)c_\nu(\beta)O_{\alpha\alpha}O_{\beta\beta} \to \overline{O_{\alpha\alpha}}^2\sum_{\nu \neq \mu}\sum_{\substack{\alpha\beta\\\alpha \neq \beta}} c_\mu(\alpha)c_\nu(\alpha)c_\mu(\beta)c_\nu(\beta)$ which we can see is consistent if this term is equal to $-\overline{O_{\alpha\alpha}}^2$. One can in-fact show simply from expanding the orthogonality condition $\sum_{\nu}\langle\psi_\mu|\psi_\nu\rangle = 0 \hspace{0.4em} | \hspace{0.4em} \mu \neq \nu$ the relation
\begin{equation}\label{eq:Average_corr}
\sum_{\nu \neq \mu}\sum_{\substack{\alpha\beta\\\alpha \neq \beta}}c_\mu(\alpha)c_\nu(\alpha)c_\mu(\beta)c_\nu(\beta) = -1 + \text{IPR}(\ket{\phi_\alpha})).
\end{equation}
where $\text{IPR}(\ket{\phi_\alpha}) := \sum_{\mu}|c_\mu(\alpha)|^4$  is the inverse partition ratio (IPR)\footnote{Notice that our definition of IPR differs here from the one used in Ref. \cite{Clos2016a} and in other works in the field of quantum chaos (e.g. \cite{Srednicki1999}) where the reciprocal quantity is defined as the IPR. Our definition in this article is more consistent with the original notion of Participation Ratio as the number of energy eigenstates or atomic orbitals involved in the initial state (see for example: 
D.J. Thouless, PHYSICS REPORTS (Section C of Physics Letters) 13,  93—142 (1974))}, which is small for systems in which our self averaging procedure is correct. Thus we find that the self-averaging assumption is consistent when applied without use of the random wave-function ansatz.

The above analysis indicates that correlations between probability amplitudes do in fact play a role, and that the common assumption that the coefficients may be treated as uncorrelated random numbers is na\"{i}ve. The illustration above is valid for generic systems with no special symmetries or correlations caused by features of the interaction, and thus the only source of these correlations is the orthonormality requirement of eigenstates. Indeed, we will see below that by including these correlations the correct scaling is obtained. 

\section{Model for Generic Non-Integrable Quantum Systems}

We now present the random matrix model from which we will base our analysis, consisting of a non-interacting 
diagonal part, and interactions modelled by a random matrix. 
Explicitly, the Hamiltonian in question is given by
\begin{eqnarray}\label{eq:Hamiltonian}
H = H_0 + V , \ \ \  H_{\alpha \beta} = f_\alpha \delta_{\alpha \beta} + h_{\alpha \beta},
\end{eqnarray}
where the diagonal matrix elements, $f_\alpha = \alpha \omega_0$, are energies equally spaced by $\omega_0$, and we choose energy units such that $\omega_0 = 1/N$, with $N$ the total number of levels.
The perturbation term is a real random Gaussian Hermitian matrix, $h$, which follows the probability distribution $P(h) \propto\exp{[-\frac{1}{4} N\Tr{h^2}/g^2]}$, such that matrix elements $h_{\alpha \beta}$ have average $\overline{h_{\alpha \beta}} = 0$, and variance  $\overline{(h_{\alpha \beta})^2} = g^2/N$ for $\alpha \neq \beta$, and $\overline{(h_{\alpha \alpha})^2} = 2g^2/N$ for diagonal elements. This is the same Hamiltonian used in the pioneering work by Deutsch \cite{Deutsch1991, Deutscha}, which captures the behaviour of a generic non-integrable quantum system in the thermodynamic limit.

We no longer restrict ourselves to observables that are diagonal in the basis of $H_0$, and thus for a generic observable $O$ we have,
\begin{equation}{\label{eq:off_diags}}
|O_{\mu\nu}|^2_{\mu \neq \nu} = \sum_{\alpha\beta\alpha^\prime\beta^\prime}c_\mu(\alpha)c_\nu(\beta)c_\mu(\alpha^\prime)c_\nu(\beta^\prime)O_{\alpha\beta}O_{\alpha^\prime\beta^\prime},
\end{equation}
where, to reiterate, we have defined $O_{\mu\nu} := \langle \psi_\mu|O|\psi_\nu\rangle$, and $O_{\alpha\beta} := \langle \phi_\alpha|O|\phi_\beta\rangle$, such that $\alpha, \beta$ labels the non-interacting basis diagonalizing $H_0$, $\{|\phi_\alpha\rangle\}$, and $\mu, \nu$ labels the interacting basis diagonalizing $H$, $\{|\psi_\mu\rangle\}$. The coefficients $c_\mu(\alpha)$ are random variables representing the eigenstates of $H$, $|\psi_\mu \rangle = c_\mu(\alpha) |\phi_\alpha\rangle$.
 
In order to obtain a functional form for the off-diagonal observable elements $|O_{\mu\nu}|^2_{\mu\neq\nu}$ we are thus interested in finding the correlation function 
\begin{equation}{\label{eq:CF}}
\langle c_\mu(\alpha)c_\nu(\beta)c_\mu(\alpha^\prime)c_\nu(\beta^\prime)\rangle_V,
\end{equation}
where the average $\langle \cdots \rangle_V$ is taken over realizations of the random Hamiltonian. We can see from the argument of the previous section that the $c_\mu(\alpha)$s are not true random variables, but have correlations due to orthogonality which must be accounted for. The probability distribution of the $c_\mu(\alpha)$ coefficients is given by
\begin{equation}\label{eq:int_dH}
P(c) = A \delta(cc^T - I)  \int
 \exp\Bigg[-\sum_{\substack{\alpha \beta \\ \alpha > \beta }}\frac{h_{\alpha\beta}^2}{2g^2/N} - \sum_{\alpha}\frac{h_{\alpha\alpha}^2}{4g^2/N} \Bigg] \Bigg( \prod_{\substack{\mu\nu\\ \mu>\nu}}\delta(\sum_{\alpha^\prime\beta^\prime}c_\mu(\alpha^\prime)H_{\alpha^\prime\beta^\prime}c_\nu(\beta^\prime)) \Bigg) \prod_{\substack{\alpha\beta \\ \alpha\geq \beta}} dh_{\alpha\beta}.
\end{equation}

In Eq. \eqref{eq:int_dH} we use the shorthand notation, $c$, to represent the matrix of $c_\mu(\alpha)$s. 
$A$ is a normalization constant, and we perform the integral over all independent entries of random Hamiltonian matrix 
elements, $h_{\alpha,\beta}$. Further, we have used $\exp[\sum_{\substack{\alpha\beta\\ \alpha\neq\beta}}h^2_{\alpha\beta}] =  \exp[2\sum_{\substack{\alpha\beta\\ \alpha>\beta}}h^2_{\alpha\beta}]$, from which we can see that, for the random matrix selected from the Gaussian Orthogonal Ensemble (GOE), the width of the distribution diagonal elements is twice that of the off-diagonal elements.
The first delta-function in $P(c)$ imposes an orthonormalization constraint whereas the last delta-function restricts the values of the $c_\mu(\alpha)$ to those of eigenstates of the Hamiltonian \eqref{eq:Hamiltonian}; the Hermiticity of $H$ implies that the latter need only run over $\mu>\nu$. 
Working with the exact probability distribution $P(c)$ is obviously very difficult. 
Studies of quantum chaotic systems \cite{Berry1977} indicate, however, that probability amplitudes behave as Gaussian distributed random variables, suggesting we may treat the $c_\mu(\alpha)$s as belonging to a Gaussian distribution with some width depending on $\mu, \alpha$. However, as we saw in Section II above, we must account for orthogonality in order to obtain a consistent result for off-diagonal matrix elements of observables. We thus look for an approximate probability distribution of the $c_\mu(\alpha)$'s of the form
\begin{equation}\label{eq:approx_prob_dist}
p(c, \Lambda) = \frac{1}{Z_p} \exp{\bigg[-\sum_{\mu\alpha}\frac{c_\mu^2(\alpha)}{2\Lambda(\mu, \alpha)}\bigg]}\prod_{\substack{\mu\nu\\\mu>\nu}} \delta(\sum_\alpha c_{\mu}(\alpha)c_{\nu}(\alpha)).
\end{equation}
In Eq. \eqref{eq:approx_prob_dist} we assume an approximation in terms of independent Gaussian variables, however, we keep the orthonormality constraint to account for correlations.
To find the functions $\Lambda(\mu,\alpha)$ that lead to an optimal description of the problem we have to  minimize the Free Energy,
\begin{equation}{\label{eq:FreeEnergy}}
F = - \int dc p(c, \Lambda) \ln{\frac{P(c)}{p(c, \Lambda)}},
\end{equation}
where we have written $\int dc$ as shorthand for an integral over all elements, $\int dc \to \prod_{\mu\alpha}\int dc_\mu(\alpha)$. The calculation of the distributions $\Lambda(\mu, \alpha)$ which fulfil this condition is performed (using a differing target probability distribution $p(c, \Lambda)$) in reference \cite{Deutscha}. We repeat this calculation in Appendix \ref{App:Lambda} for clarity. We obtain
\begin{equation}\label{eq:Lambda}
\Lambda(\mu, \alpha) = \omega_0 \frac{\Gamma / \pi}{(\omega_0\mu - \omega_0 \alpha)^2 +  \Gamma^2}.
\end{equation}
where $\Gamma = \frac{\pi g^2}{N\omega_0} $, differing by a factor of 2 from reference \cite{Deutscha} (this is corroborated below with a numerical calculation). Also required for the calculation of the correlation function (\ref{eq:CF}) is the partition function of our approximate probability distribution, which is also obtained in Appendix \ref{App:Lambda} (Eq. (\ref{eq:PartFuncFullApp})):
\begin{equation}{\label{eq:PartFuncFull}}
Z_p = (2\pi)^{N^2-N/2}\left(\prod_{\mu\alpha} (\Lambda({\mu},\alpha))^{\frac{1}{2}}\right) \left( \prod_{\substack{\mu\nu\\\mu>\nu}} \bigg(\sum_{\alpha} \Lambda(\mu, \alpha)\Lambda(\nu, \alpha) \bigg)^{-{\frac{1}{2}}} \right).
\end{equation}
In Eq. \eqref{eq:PartFuncFull} the first product is the contribution from the free Gaussian term in $p(c,\Lambda)$, whereas the second product is a result of the orthonormality condition.
\section{Calculation of Correlation Functions}
We can see from Eq. (\ref{eq:PartFuncFull}) that the final form of the partition function describing the full system is a product of all eigenvector interactions occurring in pairs. 
We are interested now in the calculation of the correlation function \eqref{eq:CF} involving a pair of random wave-functions, $c_\mu(\alpha)$ and $c_\nu(\alpha)$. For that we define the generating function
\begin{eqnarray}{\label{eq:PartFunc}}
& & G_{\mu,\nu} (\vec{\xi}_{\mu},\vec{\xi}_{\nu}) = \nonumber \\ 
& & \int \int  \exp{\bigg[-\sum_\alpha( \frac{c_\mu^2(\alpha)}{2\Lambda(\mu, \alpha)} + \frac{c_\nu^2(\alpha)}{2\Lambda(\nu, \alpha)} + \sum_\alpha(\xi_{\mu,\alpha} c_{\mu}(\alpha) + \xi_{\nu,\alpha} c_{\nu}(\alpha) ) )\bigg]}
\delta(\sum_{\alpha}c_\mu(\alpha)c_\nu(\alpha))\prod_{\alpha}dc_\mu(\alpha) dc_\nu(\alpha).
\end{eqnarray}
We will calculate correlation functions by differentiation of $G_{\mu,\nu}$ with respect to the auxiliary fields $\xi_{\mu,\alpha}$, $\xi_{\nu,\alpha}$, described for all $\alpha$ by $\vec{\xi}_\mu, \vec{\xi}_\nu$. This approach involves an implicit approximation, namely, we are assuming that correlations involving two random wavefunctions can be computed by singling out the contribution of those wavefunctions to the partition function and factoring out the rest. This approximation is well justified since it accounts for the effect of the orthonormality between $\mu$ and $\nu$, which will determine the form of the correlation function.

Eq. (\ref{eq:PartFunc}) may be evaluated as a $2N$-dimensional Gaussian integral after we express the delta-function in its Fourier form,
\begin{equation}
\delta(\sum_\alpha c_\mu(\alpha) c_\nu(\alpha))  
= \frac{1}{2 \pi} \int\exp{\bigg[i \lambda \sum_\alpha c_\mu(\alpha) c_\nu(\alpha)\bigg]}  d \lambda .
\end{equation}
We write our generating function in the form
\begin{equation}{\label{eq:GaussianForm}}
G_{\mu,\nu} (\vec{\xi}_{\mu},\vec{\xi}_{\nu}) =  \frac{1}{2\pi}\int  \int \exp{\bigg[-\frac{1}{2}\vec{x}^T\mathbf{A}\vec{x} + \vec{J}^T\vec{x}\bigg]}d^{2N}x d\lambda,
\end{equation}
where $\vec{x} = (c_\mu(1), c_\nu(1), ... , c_\mu(N), c_\nu(N))$ is a vector made up of coefficients of both relevant eigenvectors, $\mathbf{A}$ is a block diagonal matrix given by
\[
\mathbf{A} = 
\begin{bmatrix}
\frac{1}{\Lambda(\mu, 1)} & i \lambda &  &  &\\
i \lambda  &\frac{1}{\Lambda(\nu, 1)} & &  &\bigzero\\
 & \ddots & \ddots & \ddots  & &\\
\bigzero & &  & \frac{1}{\Lambda(\mu, N)} & i\lambda \\
 & & & i \lambda  & \frac{1}{\Lambda(\nu, N)}

\end{bmatrix},
\]
and $\vec{J} = (\xi_{\mu, 1}, \xi_{\nu, 1}, ..., \xi_{\mu, N}, \xi_{\nu, N})$ is the generating function for the calculation of the correlation functions.  Eq. (\ref{eq:GaussianForm}) may then be calculated exactly, as the $2N$-dimensional integral over $x$ is now in Gaussian form, and is given by
\begin{equation}
\int \exp{\bigg[-\frac{1}{2}\vec{x}^T\mathbf{A}\vec{x} + \vec{J}^T\vec{x}\bigg]}d^{2N}x  = \frac{(2\pi)^{N}}{|\mathbf{A}|^{\frac{1}{2}}} \exp{\bigg[\frac{1}{2}\vec{J}^T\mathbf{A}^{-1}\vec{J}\bigg]}.
\end{equation}
where $|\mathbf{A}|$ is the determinant of the block diagonal matrix $\mathbf{A}$, given by the product of the determinants of each $2\times 2$ block,
\begin{equation}
|\mathbf{A}| := \prod_\alpha |A_\alpha| := \prod_\alpha \bigg(\frac{1}{\Lambda(\mu, \alpha)\Lambda(\nu, \alpha)} + \lambda^2\bigg),
\end{equation}
and 
\begin{equation}
{\frac{1}{2}\vec{J}^T\mathbf{A}^{-1}\vec{J}} = -\frac{1}{2}\sum_\alpha \frac{1}{|A_\alpha|}\bigg[\frac{\xi_{\mu,\alpha}^2}{\Lambda(\nu, \alpha)} + \frac{\xi_{\nu,\alpha}^2}{\Lambda(\mu, \alpha)} - 2i\lambda \xi_{\mu,\alpha}\xi_{\nu,\alpha}\bigg].
\end{equation}
We then have,
\begin{equation}\label{eq:PartFunc2}
G_{\mu,\nu} (\vec{\xi}_{\mu},\vec{\xi}_{\nu}) = \lim_{\vec{\xi}_{\mu, \nu} \to 0} (2\pi)^{N-1}\int  \frac{1}{|\mathbf{A}|^{\frac{1}{2}}} \exp{\Bigg[-\frac{1}{2}\sum_\alpha \frac{1}{|A_\alpha|}\bigg[ \frac{\xi_{\mu,\alpha}^2}{\Lambda(\nu, \alpha)} + \frac{\xi_{\nu,\alpha}^2}{\Lambda(\mu, \alpha)} - 2i\lambda \xi_{\mu,\alpha}\xi_{\nu,\alpha}\bigg]\Bigg]}d\lambda,
\end{equation}
which we write as
\begin{equation}\label{eq:PartFunc3}
G_{\mu,\nu} (\vec{\xi}_{\mu},\vec{\xi}_{\nu}) = \lim_{\vec{\xi}_{\mu, \nu} \to 0} (2\pi)^{N-1}\int  \left(\prod_{\alpha}\bigg(\frac{1}{\Lambda(\mu, \alpha)\Lambda(\nu, \alpha)} + \lambda^2\bigg)^{-\frac{1}{2}} g(\xi_{\mu\alpha}, \xi_{\nu\alpha})\right)d\lambda,
\end{equation}
where
\begin{equation} \label{def.g}
g(\xi_{\mu,\alpha}, \xi_{\nu,\alpha}) = \exp{\bigg[ \frac{1}{2}\frac{\xi_{\mu,\alpha}^2\Lambda(\mu, \alpha) + \xi_{\nu,\alpha}^2\Lambda(\nu, \alpha) - 2i\lambda \xi_{\mu,\alpha} \xi_{\nu,\alpha}\Lambda(\mu, \alpha)\Lambda(\nu, \alpha)}{1 + \lambda^2 \Lambda(\mu, \alpha)\Lambda(\nu, \alpha)} \bigg]}.
\end{equation}
Now, we can rewrite the integrand in Eq. (\ref{eq:PartFunc3}) as 
\begin{equation}
(2\pi)^{N-1} \left( \prod_{\alpha} \big(\Lambda(\mu, \alpha)\Lambda(\nu, \alpha)\big)^{\frac{1}{2}}\right)\left( \prod_{\alpha^\prime}\big(1 + \lambda^2\Lambda(\mu, \alpha^\prime)\Lambda(\nu, \alpha^\prime)\big)^{-\frac{1}{2}}\right)\left( \prod_{\alpha^{\prime\prime}}g(\xi_{\mu\alpha^{\prime\prime}}, \xi_{\nu\alpha^{\prime\prime}})\right).
\end{equation}
 Then, as $\ln{(1+x)} \approx x$ for small $x$, in the high $N$ limit we have 
\begin{equation}
 \begin{split}
\prod_{\alpha}\big(1 + \lambda^2\Lambda(\mu, \alpha)\Lambda(\nu, \alpha)\big)^{-\frac{1}{2}} & = \exp\bigg[{-\frac{1}{2}\sum_{\alpha}\ln{\Big(1 + \lambda^2\Lambda(\mu, \alpha)\Lambda(\nu, \alpha)\Big)}}\bigg] \\ & \approx \exp{\bigg[-\frac{1}{2}\sum_{\alpha}\lambda^2\Lambda(\mu, \alpha)\Lambda(\nu, \alpha)\bigg]}.
 \end{split}
\end{equation}
Thus, we obtain for the generating function
\begin{equation}{\label{eq:PF}}
G_{\mu,\nu} (\vec{\xi}_{\mu},\vec{\xi}_{\nu}) = \lim_{\vec{\xi}_{\mu, \nu} \to 0} (2\pi)^{N-1}\left(\prod_\alpha \Lambda(\mu, \alpha)\Lambda(\nu, \alpha)\right)^{\frac{1}{2}}\int  \exp{\bigg[-\frac{1}{2}\sum_{\alpha}\lambda^2\Lambda(\mu, \alpha)\Lambda(\nu, \alpha)\bigg]}\left(\prod_{\alpha^\prime} g(\xi_{\mu\alpha^\prime}, \xi_{\nu\alpha^\prime})\right) d\lambda.
\end{equation}
The generation function can be checked to yield the correct $\vec{\xi}_{\mu} = 0$, $\vec{\xi}_{\nu} = 0$ limit, 
\begin{equation}\label{eq:Zapprox}
G_{\mu\nu}(0,0) 
= 
(2\pi)^{N-\frac{1}{2}}\bigg(\frac{\prod_\alpha \Lambda(\mu, \alpha)\Lambda(\nu, \alpha)}{\sum_\beta \Lambda(\mu, \beta) \Lambda(\nu, \beta)}\bigg)^\frac{1}{2}.
\end{equation}
Taking the product over all pairs of eigenvectors $\mu, \nu$ of the 2-eigenvector partition function of Eq. (\ref{eq:Zapprox}) we recover the interacting part of the partition function of the previous section, Eq. (\ref{eq:PartFuncFull}). 

We can proceed now and simplify the generating function by simplifying Eq. \eqref{def.g} in the limit $\Gamma/\omega_0 \gg 1$. For this, we first notice that, due to the Gaussian term in Eq. \eqref{eq:PF}, the integration variable $\lambda$ is restricted to take values such that,
\begin{equation}
\lambda^2 \sum_{\alpha} \Lambda(\mu,\alpha) \Lambda(\nu,\alpha) \lesssim 1 .
\end{equation}
Since the term $\sum_{\alpha} \Lambda(\mu,\alpha) \Lambda(\nu,\alpha)$ is of order $(\Gamma/\omega_0)^{-1}$, this implies that $\lambda \approx (\Gamma/\omega_0)^{1/2}$. On the other hand, in Eq. \eqref{def.g}, we find in the denominator the term
$\lambda^2 \Lambda(\mu,\alpha) \Lambda(\nu,\alpha)$. Since the product  $\Lambda(\mu,\alpha) \Lambda(\nu,\alpha)$ takes values of the order of $(\Gamma/\omega_0)^{-2}$, we find that
\begin{equation}
\lambda^2 \Lambda(\mu,\alpha) \Lambda(\nu,\alpha) = {\cal O} \left( \frac{\Gamma}{\omega_0} \right)^{-1} \ll 1 .
\end{equation}
Using this approximation and carrying out the integration over $\lambda$ we arrive at the following form for the generating function,
\begin{equation}\label{final.G}
G_{\mu,\nu} (\vec{\xi}_{\mu},\vec{\xi}_{\nu}) \propto 
\exp{ \bigg[ 
\frac{1}{2} \sum_\alpha \xi^2_{\mu,\alpha} \Lambda(\mu,\alpha) +
\frac{1}{2} \sum_\alpha \xi^2_{\nu,\alpha} \Lambda(\nu,\alpha) -
\frac{1}{2} \sum_{\alpha,\beta} 
\xi_{\mu,\alpha} \xi_{\mu,\beta} \xi_{\nu,\alpha} \xi_{\nu,\beta} 
	\frac{\Lambda(\mu,\alpha) \Lambda(\mu,\beta) \Lambda(\nu,\alpha) \Lambda(\nu,\beta)}{\sum_{\alpha'} \Lambda(\mu,\alpha') \Lambda(\nu,\alpha')}  
\bigg]},
\end{equation}
where we have ignored the non-interacting factors, which are irrelevant for the calculation of the correlation functions. Eq. \eqref{final.G} is the basis of a self-consistent description of matrix elements in terms of random wave-functions.

We apply our result for the correlation function of interest (see Eq. (\ref{eq:off_diags})),
\begin{equation}{\label{eq:4ptcorr}}
\langle c_\mu(\alpha) c_\nu(\beta) c_\mu(\alpha^\prime) c_\nu(\beta^\prime) \rangle_V = \frac{1}{G_{\mu\nu}} \partial_{\xi_{\mu,\alpha}} \partial_{\xi_{\nu,\beta}} \partial_{\xi_{\mu,\alpha^\prime}} \partial_{\xi_{\nu,\beta^\prime}} G_{\mu\nu} {\bigg |}_{\xi_{\mu,\alpha}=0,\xi_{\nu,\alpha}=0}.
\end{equation}
After calculating the derivatives of our simplified generating function \eqref{final.G} we obtain,
\begin{equation}\label{eq:Corr1offdiag}
\begin{split}
&\langle c_\mu(\alpha) c_\nu(\beta) c_\mu(\alpha^\prime) c_\nu(\beta^\prime) \rangle_V = \Lambda(\mu, \alpha)\Lambda(\nu, \beta)\delta_{\alpha\alpha^\prime}\delta_{\beta\beta^\prime} \\& - \frac{\Lambda(\mu, \alpha)\Lambda(\nu, \alpha)\Lambda(\mu, \beta)\Lambda(\nu, \beta)\delta_{\alpha\beta^\prime}\delta_{\beta\alpha^\prime}}{\sum_n \Lambda(\mu, n)\Lambda(\nu, n)}  - \frac{\Lambda(\mu, \alpha)\Lambda(\nu, \alpha)\Lambda(\mu, \alpha^\prime)\Lambda(\nu, \alpha^\prime)\delta_{\alpha\beta}\delta_{\alpha^\prime\beta^\prime}}{\sum_n \Lambda(\mu, n)\Lambda(\nu, n)}.
\end{split}
\end{equation}
In the last equation, the second and third terms in the right-hand side arise solely due to the interactions between random wave-functions that are induced by the orthonormality condition.

For an observable that is diagonal in the basis of $H_0$ we only need to consider the values $\alpha = \beta$ and $\alpha' = \beta'$. The relevant correlation function is then of the simpler form
\begin{equation}\label{eq:Corr1offdiagLimit}
\langle c_\mu(\alpha) c_\nu(\alpha) c_\mu(\beta) c_\nu(\beta) \rangle_V = \Lambda(\mu, \alpha)\Lambda(\nu, \beta)\delta_{\alpha\beta} - \frac{\Lambda(\mu, \beta)\Lambda(\nu, \alpha)\Lambda(\mu, \alpha)\Lambda(\nu, \beta)(1 + \delta_{\alpha\beta})}{\sum_n \Lambda(\mu, n)\Lambda(\nu, n)}.
\end{equation}
Eq. (\ref{eq:Corr1offdiagLimit}) is one of the most important results of this work. Note that the first term in the r.h.s. of this equation is the contribution one obtains by ignoring the interaction between random wave-functions, whereas the second term arises solely due to those interactions. It is thus necessary to understand whether the corrections induced by interactions are relevant or, on the contrary, can be neglected to leading order (as assumed in many previous works). 
For this we first notice that
\begin{equation}
\Lambda(\mu,\alpha)|_{E_\mu \approx E_\alpha} \approx \frac{\omega_0}{\Gamma},
\end{equation}
where the ratio $\omega_0/\Gamma \ll 1$, since it corresponds to the inverse number of states in the energy window defined by $\Gamma$. We find the following scaling
\begin{eqnarray}\label{eq:scaling_d}
\Lambda(\mu,\alpha)\Lambda(\nu,\alpha) \hspace{0.5cm} &\to& 
{\cal O}\left(\frac{\omega_0}{\Gamma}\right)^2 , \\
\label{eq:scaling_nd}
\frac{\Lambda(\mu, \beta)\Lambda(\nu, \alpha)\Lambda(\mu, \alpha)\Lambda(\nu, \beta)}{\sum_n \Lambda(\mu, n)\Lambda(\nu, n)}
&\to& {\cal O}\left(\frac{\omega_0}{\Gamma}\right)^3.
\end{eqnarray}
We could feel tempted to simply ignore the correlation term in Eq. (\ref{eq:Corr1offdiagLimit}), since it is of higher order in the small parameter $\omega_0/\Gamma$. 
Neglecting the correlation term is a valid approximation in the case $\alpha = \beta$, since we find that the leading term contribution is given by Eq. (\ref{eq:scaling_d}). 
On the contrary, for non-diagonal terms ($\alpha \neq \beta$), 
the lowest order contribution is given by Eq. (\ref{eq:scaling_nd}) and it is of order  ${\cal O}\left(\frac{\omega_0}{\Gamma}\right)^3$. However, there are of order $\Gamma/\omega_0$ more non-diagonal than diagonal terms. Whenever we use the correlation function 
Eq. (\ref{eq:Corr1offdiagLimit}) to calculate the expectation value of an observable, we will need to sum over indices $\alpha, \beta$. Thus we expect that the contribution of ${\cal O}\left(\frac{\Gamma}{\omega_0}\right)$ non-diagonal terms each contributing an amount of order ${\cal O}\left(\frac{\omega_0}{\Gamma}\right)^3$ will yield finally a contribution or order ${\cal O}\left(\frac{\omega_0}{\Gamma}\right)^2$, which is thus comparable to the contribution from the diagonal terms. We conclude that both terms in the r.h.s. of Eq. (\ref{eq:Corr1offdiagLimit}) are equally relevant.

The reasoning above also explains discrepancies that one may find when, for example, verifying the orthonormality sum rule with Eq. (\ref{eq:Corr1offdiagLimit}). Explicitly, orthnormality implies that,
\begin{equation}
\sum_{\nu} \langle c_\mu(\alpha) c_\nu(\alpha) c_\mu(\beta) c_\nu(\beta) \rangle =
 \Lambda(\mu, \alpha) \delta_{\alpha,\beta}.
\end{equation}
However, Eq. (\ref{eq:Corr1offdiagLimit}) yields,
\begin{equation}
\sum_{\nu} \langle c_\mu(\alpha) c_\nu(\alpha) c_\mu(\beta) c_\nu(\beta) \rangle_V =
 \Lambda(\mu, \alpha) \delta_{\alpha,\beta} 
 + {\cal O}\left( \frac{\omega_0}{\Gamma} \right)^2 .
\label{eq:CheckSumRule}
\end{equation}
The correction of order ${\cal O}\left(\frac{\omega_0}{\Gamma} \right)^2$ 
can be ignored, since the leading contribution to the diagonal term is $\Gamma(\mu,\alpha)$, which is of order 
${\cal O}\left(\frac{\omega_0}{\Gamma} \right)$.
The attentive reader may find a contradiction in neglecting terms that are one order lower in $\omega_0 / \Gamma$ in Eq. (\ref{eq:CheckSumRule}), while keeping the second term in the r.h.s. of Eq. (\ref{eq:Corr1offdiagLimit}). However, we recall that in the latter case, we have to sum over a large number of low-order non-diagonal corrections, and thus both Eqs. 
(\ref{eq:scaling_d}) and (\ref{eq:scaling_nd}) may lead to contributions of the same order when calculating  matrix elements of observables.

We also stress here that whilst the derivation of the Lorentzian form of 
$\Lambda(\mu, \alpha)$ is perturbative, and thus only accurate for small couplings, our result of Eq. (\ref{eq:Corr1offdiag}) is more general and relies only on the condition that the wave-function is spread over many non-interacting states. 
For example, a system with a Gaussian form $\Lambda(\mu, \alpha)$ could be described by the approximate distribution \eqref{eq:approx_prob_dist}, and yet lead to the same form for the random wave-function correlations.

\section{Calculation of Off-Diagonal Matrix Elements}
  
We can now use the functional form for $\langle c_\mu(\alpha)c_\nu(\beta)c_\mu(\alpha^\prime)c_\nu(\beta^\prime)\rangle_V$ developed in the previous section to calculate a generic form for $|O_{\mu\nu}|^2$. We have
\begin{equation}{\label{eq:Off_diags2}}
|O_{\mu\nu}|^2_{\mu \neq \nu} = \sum_{\alpha\beta\alpha^\prime\beta^\prime}c_\mu(\alpha)c_\nu(\beta)c_\mu(\alpha^\prime)c_\nu(\beta^\prime)O_{\alpha\beta}O_{\alpha^\prime\beta^\prime}.
\end{equation}
Now, assuming self-averaging, we can replace $c_\mu(\alpha)c_\nu(\beta)c_\mu(\alpha^\prime)c_\nu(\beta^\prime) \to \langle c_\mu(\alpha)c_\nu(\beta)c_\mu(\alpha^\prime)c_\nu(\beta^\prime) \rangle_V$. Then, using our expression for the correlation function, Eq. (\ref{eq:Corr1offdiag}) we can write
\begin{equation}\label{eq:O_full}
\begin{split}
|O_{\mu\nu}|^2_{\mu \neq \nu} &= \sum_{\alpha\beta\alpha^\prime\beta^\prime}\bigg[ \Lambda(\mu, \alpha)\Lambda(\nu, \beta)\delta_{\alpha\alpha^\prime}\delta_{\beta\beta^\prime}  - \frac{\Lambda(\mu, \alpha)\Lambda(\nu, \alpha)\Lambda(\mu, \beta)\Lambda(\nu, \beta)\delta_{\alpha\beta^\prime}\delta_{\beta\alpha^\prime}}{\sum_n \Lambda(\mu, n)\Lambda(\nu, n)} \\& - \frac{\Lambda(\mu, \alpha)\Lambda(\nu, \alpha)\Lambda(\mu, \alpha^\prime)\Lambda(\nu, \alpha^\prime)\delta_{\alpha\beta}\delta_{\alpha^\prime\beta^\prime}}{\sum_n \Lambda(\mu, n)\Lambda(\nu, n)}\bigg]O_{\alpha\beta}O_{\alpha^\prime\beta^\prime}.
\end{split}
\end{equation}
If we once more briefly focus on those observables that are diagonal in the $H_0$ eigenbasis, this becomes
\begin{equation}
|O_{\mu\nu}|^2_{\mu \neq \nu}  = \sum_{\alpha}\Lambda(\mu, \alpha)\Lambda(\nu, \alpha)O_{\alpha\alpha}^2 - \frac{\sum_{\alpha}\Lambda(\mu, \alpha)\Lambda(\nu, \alpha)O_{\alpha\alpha}\sum_{\beta}\Lambda(\mu, \beta)\Lambda(\nu, \beta)O_{\beta\beta}}{\sum_n \Lambda(\mu, n)\Lambda(\nu, n)} - \frac{\sum_{\alpha}\Lambda^2(\mu, \alpha)\Lambda^2(\nu, \alpha)O_{\alpha\alpha}^2}{\sum_n \Lambda(\mu, n)\Lambda(\nu, n)}.
\end{equation}
Again, we find that a non-negligible contribution arises from the random wave-function correlations. To further approximate this expression we define the average
\begin{equation}
\overline{[O_{\alpha\alpha}]}_{\overline{\mu}} := \sum_{\alpha}\Lambda(\overline{\mu}, \alpha)O_{\alpha\alpha},
\end{equation}
where $\overline{\mu} := (\mu + \nu)/2$, which one may observe is essentially a microcanonical average centered on the energy $E_{\overline{\mu}}$. A further self-averaging approximation allows this microcanonical average to be removed from the summation.
\begin{equation}
\label{eq:offdiag}
|O_{\mu\nu}|^2_{\mu \neq \nu}  = \bigg(\overline{[O_{\alpha\alpha}^2]}_{\overline{\mu}} - \overline{[O_{\alpha\alpha}]}_{\overline{\mu}}^2 \bigg)\sum_{\alpha}\Lambda(\mu, \alpha)\Lambda(\nu, \alpha) - \overline{[O^2_{\alpha\alpha}]}_{\overline{\mu}}\frac{\sum_{\alpha}\Lambda^2(\mu, \alpha)\Lambda^2(\nu, \alpha)}{\sum_n \Lambda(\mu, n)\Lambda(\nu, n)}.
\end{equation}
Eq. \eqref{eq:offdiag} is one of the most important results of this work. Note that the result is now free from the pathology that we found when approximating many-body wave-functions by independent random numbers in Eq. \eqref{eq:URN_1}. Our final expression has a similar form, however correlations induce a second term that appears as a result of the orthonormality condition. Finally we note that the overall dependence of $|O_{\mu,\nu}|^2$ on the energies $E_\mu,E_\nu$ agrees with the ETH ansatz for off-diagonal matrix elements in Eq. \eqref{eq:offdiagETH}.

We then take the continuum limit, substituting $\sum_\alpha \to \int\frac{dE_\alpha}{\omega_0}$, and thereby obtain
\begin{equation}{\label{eq:Oint}}
\begin{split}
|O_{\mu\nu}|^2_{\mu \neq \nu} &= \bigg(\overline{[O_{\alpha\alpha}^2]}_{\overline{\mu}} - \overline{[O_{\alpha\alpha}]}_{\overline{\mu}}^2 \bigg) \int \frac{dE_\alpha}{\omega_0}\Lambda(\mu, \alpha)\Lambda(\nu, \alpha)\\& - \overline{[O^2_{\alpha\alpha}]}_{\overline{\mu}}\int\frac{dE_\alpha}{\omega_0}\Lambda(\mu, \alpha)^2\Lambda(\nu, \alpha)^2\bigg( \int \frac{dE_n}{\omega_0}\Lambda(\mu, n)\Lambda(\nu, n)\bigg)^{-1}.
\end{split}
\end{equation}

Whilst the second term in Eq. (\ref{eq:Oint}) is analytically obtainable, we may observe that this term is $\propto \omega_0^2$, and thus within our approximation is correctly ignored. We then see, as the convolution of two Lorentzian functions of widths $\Gamma_1$ and $\Gamma_2$ is simply a Lorentzian of width $\Gamma_1 + \Gamma_2$, that the functional form for a diagonal observable is
\begin{equation}{\label{eq:O_fin}}
|O_{\mu\nu}|^2_{\mu \neq \nu} = \overline{[\Delta O_{\alpha\alpha}^2]}_{\overline{\mu}} \frac{\omega_0 2\Gamma / \pi}{(E_\mu - E_\nu)^2 + (2\Gamma)^2},
\end{equation}
where $\overline{[\Delta O_{\alpha\alpha}^2]}_{\overline{\mu}} := \overline{[O_{\alpha\alpha}^2]}_{\overline{\mu}} - \overline{[O_{\alpha\alpha}]}_{\overline{\mu}}^2$. We see here that, to first order in $\omega_0$, the off diagonal elements of a generic observable that is diagonal in $H_0$ are described by a Lorentzian of width $2\Gamma$. For more general observables one simply uses the known structure in the non-interacting basis, as we will see below. This result corroborates the relation between the variances of diagonal and off-diagonal elements obtained in reference \cite{DAlessio2016}, and observed numerically in\cite{Mondaini2017, Dymarsky2017}, showing that they differ by a factor of two. One can see that the width of the distribution of diagonal elements is the same as that of the wave-function, $\Gamma$, from Eq. (\ref{diagonal}).

Returning to our original argument indicating the failure of the random wave-function ansatz, we may double check the consistency of the above RMT approach by repeating the calculation of $\sum_{\nu} |O_{\mu\nu}|^2_{\mu\neq\nu}$ using Eq. (\ref{eq:O_fin}). This is obtained by replacing $\sum_{\nu\neq\mu} \to \int dE_\nu / \omega_0$ (the correction due to the $\mu = \nu$ term is $\propto \omega_0^2$ and thus ignored)
\begin{equation}
\begin{split}
 \int \frac{dE_\nu}{\omega_0} |O_{\mu\nu}|^2_{\mu\neq\nu} & = \int \frac{dE_\nu}{\omega_0} \overline{[\Delta O_{\alpha\alpha}^2]}_{\overline{\mu}}\frac{\omega_0 2\Gamma / \pi}{(E_\mu - E_\nu)^2 + (2\Gamma)^2} \\ &  = \overline{[\Delta O_{\alpha\alpha}^2]}_{\overline{\mu}},
\end{split}
\end{equation}
as expected. Thus the RMT approach, including correlations due to orthogonality, leads to a correct normalization of the matrix elements of observables.

We note here that the result from the RMT approach tells us more about the source of this scaling than we obtained from our previous discussion. Eq. (\ref{eq:Average_corr}) tells us that the sum over all off diagonal eigenstates contributes this scaling factor, but gives us no information about the contribution of any individual eigenstate. We can see from the RMT result of Eq. (\ref{eq:O_fin}) that the scaling by $\overline{[\Delta O_{\alpha\alpha}^2]}_{\overline{\mu}}$ occurs on the level of each individual eigenstate, and not simply on average.

\section{Comparison to Numerical Random Matrix Model}
\begin{figure}
    \includegraphics[width=0.95\textwidth]{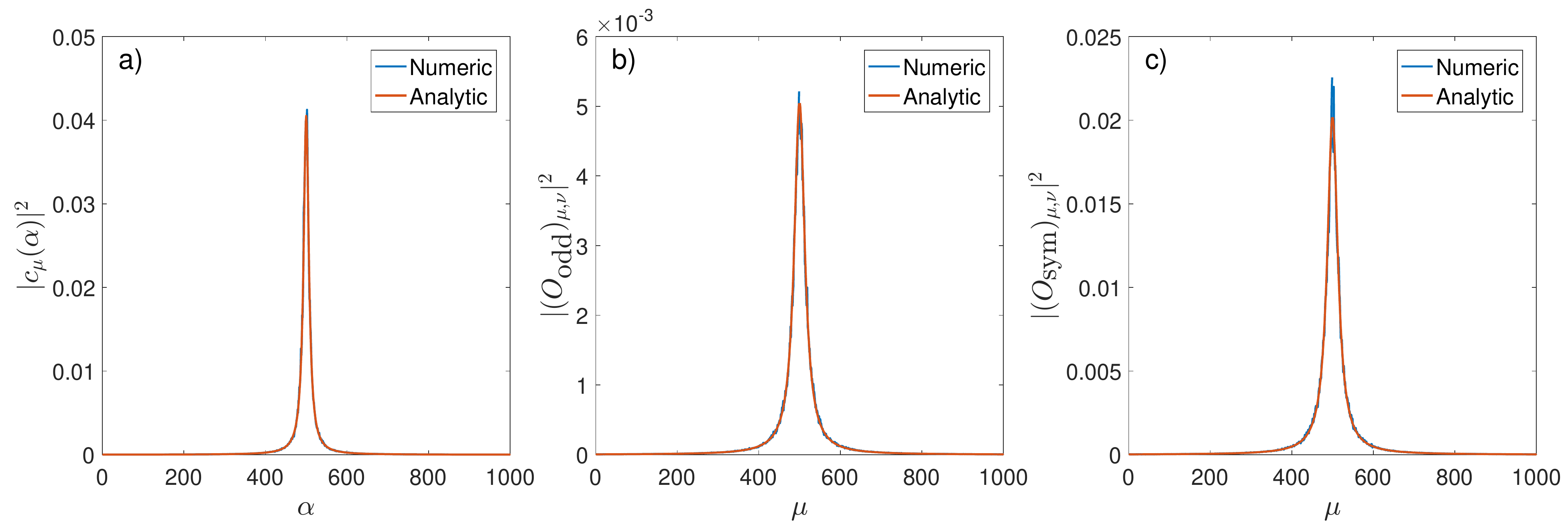}
    \caption{Numerical comparison to analytic results for $N = 1000$, $g = 0.05,$ average of 500 realizations of $H$. a) Shows the eigenstate $\mu = 500$, analytic result given by (\ref{eq:Lambda}) b),c) Show the off-diagonal distributions ($\mu = \nu$ points are excluded) for $O_{\rm odd}$ and $O_{\rm sym}$ respectively analytic result given by Eq. (\ref{eq:O_fin}). }
    \label{fig:RMT_Numerical}
\end{figure}

To check the results above we first compare them to a numerical random matrix model by diagonalizing Eq. (\ref{eq:Hamiltonian}) and calculating the off-diagonal distribution for the matrix elements of example observables. We choose our observables, $O_{\rm odd}$ and $O_{\rm sym}$, to be defined such that in the non-interacting basis $\{|\phi_\alpha \rangle \}$ all off-diagonal elements are zero, and the diagonal elements are given by
\begin{equation}
(O_{\rm odd})_{\alpha\alpha}  =\begin{cases}
    1, & \text{if $\alpha =$ odd}\\
    0, & \text{if $\alpha =$ even},
  \end{cases}
\end{equation}
and
\begin{equation}
(O_{\rm sym})_{\alpha\alpha}  =\begin{cases}
    1, & \text{if $\alpha =$ odd}\\
    -1, & \text{if $\alpha =$ even}.
  \end{cases}
\end{equation}
These observables are chosen as they have similar structure to realistic spin-observables, as well as having different $\overline{[\Delta O_{\alpha\alpha}^2]}_{\overline{\mu}}$ values such that the scaling may be adequately demonstrated. For simplicity we choose diagonal examples here, though the RMT method developed above can easily account for non-diagonal observables, as we will see below for a spin-chain system. To obtain the observable distributions we find the average distribution over many realizations of the Hamiltonian, Eq. (\ref{eq:Hamiltonian}), which is essentially the mathematical procedure to find the probability distribution in Eq. (\ref{eq:int_dH}). Examples of the overlap of the RMT prediction are shown in Fig. (\ref{fig:RMT_Numerical}). Here we see a very good agreement between the analytic predictions of Eq. (\ref{eq:Lambda}) (Fig. (\ref{fig:RMT_Numerical}a)) and Eq. (\ref{eq:O_fin}) (Figs. (\ref{fig:RMT_Numerical}b), (\ref{fig:RMT_Numerical}c)) and the exact numerical results. 

As shown in Fig. (\ref{fig:RMT_Numerical}), the scaling of each observable $O_{\rm odd}$ and $O_{\rm sym}$ are different, and we can see here that the analytic prediction of an observable dependent rescaling is true to the numerics. We note that for couplings of $g \gtrapprox 0.2$ our analytic treatment is no longer a good approximation. This corresponds to the bulk eigenstates having significant value at the edges of the spectrum, and thus our assumptions made obtaining a functional form for $\Lambda(\mu, \alpha)$ (Appendix \ref{App:Lambda}) are not good for such coupling strengths.

Before making comparison to realistic systems, we comment here on an essential ingredient to the derivation of our analytic results: the self-averaging procedure. This property of random matrices is commonly assumed\cite{DAlessio2016,Muller2015}, and whilst not rigorously proven, has been an invaluable tool in the descriptive power of RMT - indeed, RMT has seen much success in describing interacting spin systems\cite{Borgonovi2016}, and atomic and nuclear physics\cite{Brody1981, Camarda1983}. Further, the analysis of random matrices based on the above assumptions already makes up much of the basis of our understanding of the ETH\cite{DAlessio2016}, which has seen repeated numerical verifications in non-integrable models. One can write the essential assumption as, for example
\begin{equation}
|O_{\mu\nu}|^2 = \langle |O_{\mu\nu}|^2  \rangle_V = \sum_{\alpha\beta\alpha^\prime\beta^\prime}\langle c_\mu(\alpha)c_\nu(\beta)c_\mu(\alpha^\prime)c_\nu(\beta^\prime)\rangle_V O_{\alpha\beta}O_{\alpha^\prime\beta^\prime},
\end{equation}
such that the observable matrix elements are taken to be equal to their ensemble average. Note that $O_{\alpha\beta}$ are not averaged quantities, as they do not depend on the random perturbation. Our work is by no means a rigorous proof of this property, however the success of the analytic results when compared to the exact numerics may be seen as further evidence of the self-averaging property of random matrices.

\section{Comparison to Exact Diagonalization of Spin-Chain}

We now perform a comparison of the theory from random matrices outlined above to a more physical system. We choose a 1D spin chain, with a Hamiltonian of the form
\begin{equation}\label{eq:Hamil}
H = H_S + H_B + H_{SB}.
\end{equation}
We reiterate here that the Lorentzian functional form for the wave-function distribution $\Lambda(\mu, \alpha)$ is obtained in the perturbative regime (see Appendix \ref{App:Lambda}), and thus we only expect good agreement with our RMT result when the interaction Hamiltonian $H_{SB}$ is small. However, the theory developed above for correlation functions, and thus the application to observable distributions, is more general. Previous numerical studies have shown that in the high coupling limit one observes a Gaussian wave-function distribution\cite{Mondaini2017, Santos2012, Njema1996, Flambaum1998, Atlas2017}, and thus for high coupling strengths we do not expect a good overlap with the developed RMT results, however the basic phenomenology should remain unchanged. 

For our model, the system Hamiltonian $H_S$ is simply given by a spin in perpendicular fields, $B_x$ and $B_z$,
\begin{equation}
H_S = B_{z}^{(S)}\sigma_z^{(N_S)} + B_{x}^{(S)}\sigma_x^{(N_S)},
\end{equation}
where $N_S$ labels the position of the system in the chain, between 1 and $N$. The bath Hamiltonian is a spin-chain with nearest-neighbour Ising interactions in both $B_z$ and $B_x$ fields,
\begin{equation}
H_B = \sum_{n \neq N_S}( B_{z}^{(B)}\sigma_z^{(n)} + B_{x}^{(B)}\sigma_x^{(n)}) + \sum_{n \neq N_S, N_S - 1}J_B(\sigma_z^{(n)}\sigma_z^{(n+1)} + \sigma_+^{(n)}\sigma_-^{(n+1)} + \sigma_-^{(n)}\sigma_+^{(n+1)} ).
\end{equation}
The interaction part of the Hamiltonian $H_{SB}$ is given by
\begin{equation}
H_{SB} = J_I(\sigma_+^{(N_S)}\sigma_-^{(N_S\pm 1)} + \sigma_-^{(N_S)}\sigma_+^{(N_S\pm 1)}) + J_z\sigma_z^{(N_S)}\sigma_z^{(N_S\pm 1)}.
\end{equation}
Thus we have $H_0 = H_S + H_B$, and $H_I = H_{SB}$. For the analysis below we compare various limits of this system, to show where our assumptions made above do and do not hold. Each limit is non-integrable, and expected to thermalize. 

We focus here on two cases: a homogeneous chain, and the case of a weakly coupled impurity. It is the latter for which we expect the RMT description to work best, as it is here that the assumption that the density of states does not change over the coupling width is valid. It is this assumption that allows us to treat the interaction Hamiltonian as a full random matrix in Eq. (\ref{eq:Hamiltonian}). Should the density of states significantly change over the relevant coupling width, then a random matrix with some bandwidth would be required. 

Initially, for the impurity case, we set $J_B = B_{z}^{(S)} =  B_{z}^{(B)} =  B_{x}^{(B)} = 1$, $J_z = B_{x}^{(S)} = 0$, and vary $J_I$. The second limit we study is when $B_{x}^{(S)} = 1$ and $J_z = J_I$, with the chain thus being truly homogeneous when $J_z = J_I = 1$. We calculate the off-diagonal matrix elements of system observables for varying system sizes from $N = 8$ to $N = 13$. We set the system position to be $N_S = 5$ throughout. 

To test the RMT prediction for the observable and wave-function distributions we calculate these distributions directly using exact diagonalization and perform a fit to the distribution to find the observed width $\Gamma_\text{Fit}$. This is then compared to the expected width from a random matrix framework, $\Gamma_{\text{RM}}$, which we discuss below. To perform the fit we first smooth the ED result by applying a Lorentzian mask over each point such that, for smoothed eigenstates we have
\begin{equation}
\label{eq:F}
F_\mu(E_\alpha) = \sum_{\alpha}|c_\mu(\alpha)|^2 \delta_\epsilon(E_\alpha - E),
\end{equation}
where $\delta_\epsilon(E_\alpha - E) = \epsilon \pi^{-1} /[(E_\nu - E) + \epsilon^2]$. 
This function is related to the strength function introduced in quantum chaos theory \cite{Borgonovi2016}.
Similarly, for an observable $O$
\begin{equation}
\label{eq:SO}
S_O(E_\mu, E_\nu) = \sum_{\nu}|O_{\mu\nu}|^2 \delta_\epsilon(E_\nu - E).
\end{equation}
We perform a three variable (central energy, peak width $\Gamma$, and peak height) fit to a Lorentzian to find the $\Gamma_\text{Fit}$. The values for $\Gamma_\text{Fit}$ can then be compared to $\Gamma = \Gamma(W_0)$ found from the interaction Hamiltonian using the method outlined below.

\subsection{Computation of RMT Width}
For comparison of our RMT description to the ED calculation, we must be able to calculate an estimate for $\Gamma$ from the random matrix perspective. This can be obtained from the Hamiltonian, as for the random matrix we have $\Gamma_{\text{RM}} = \pi g^2 / N\omega_0$, and $g/\sqrt{N}$, which may be found by the average value of the random interaction Hamiltonian. Relating this to a physical system must be done with some care, however, as the average value should not be taken over the entire Hamiltonian, but over some energy width $W$, as discussed below. We can write $\Gamma_{\text{RM}}$, for a random matrix, as
\begin{equation}\label{eq:GammaTr1}
\Gamma_{\text{RM}} = \pi \frac{\Tr \{H_I^\dagger H_I\}}{N^2}D(E),
\end{equation}
where $D(E) = 1/\omega_0$ is the density of states. In this form we can see more easily the relation to a real Hamiltonian. However we must treat the above expression carefully, as the association $g^2 / N \Leftrightarrow \Tr \{H_I^\dagger H_I\} / N^2$ must be made with proper consideration of the physical relationship between the interaction Hamiltonian and a random matrix. To reiterate, the physical grounds for using a random interaction Hamiltonian here rely on the fact that for generic non-integrable systems the interaction Hamiltonian, when expressed in the basis of eigenstates of the non-interacting Hamiltonian, resembles a banded random matrix with some width $W_{BW}$. We can use a full random matrix for the low coupling limit as the density of states, which dictates the band width, does not change much over the width of the coupling energy $\Gamma$. Thus, there are two caveats to be considered in implementing Eq. (\ref{eq:GammaTr1}): $H_I$ must be expressed in the basis of $H_0$, and the trace must be taken over a finite width $W_0 < W_{BW}$. 
\begin{figure}
    \includegraphics[width=0.95\textwidth]{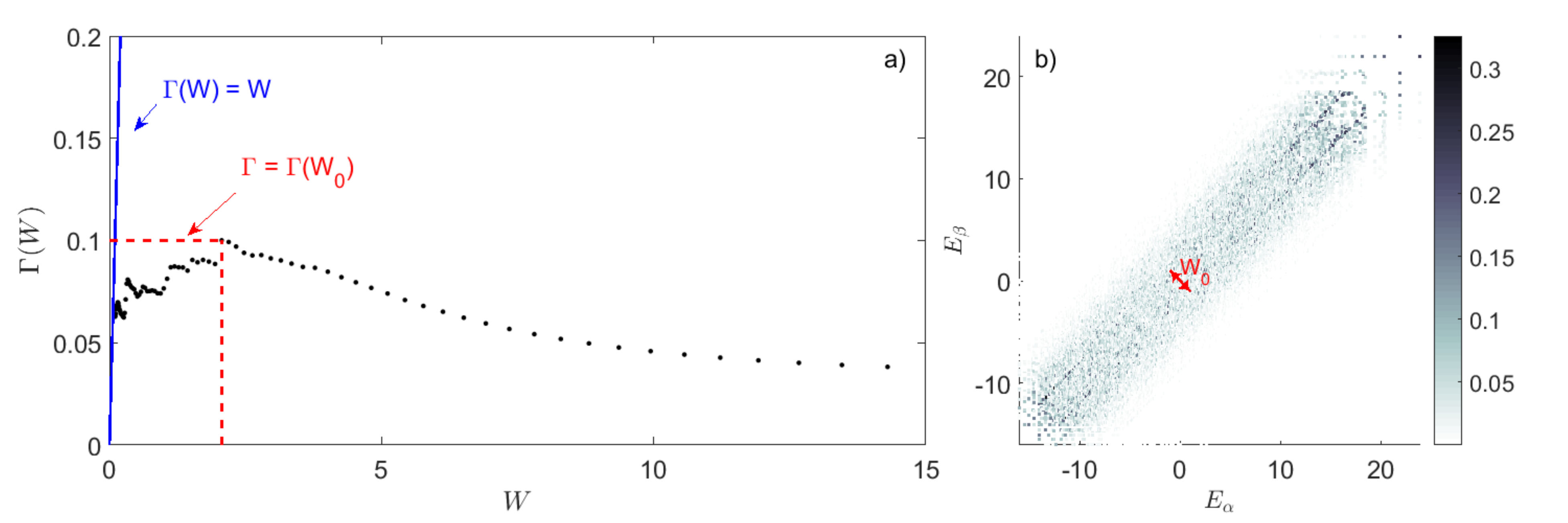}
    \caption{a) Plot of $\Gamma(W)$ (Eq. (\ref{eq:Gamma_W})) for varying values of $W$ (black dots). Approximate $\Gamma = \Gamma(W_0)$ shown by red dashed line. We see that the plateau region indeed extends well past the $\Gamma(W) = W$ line (blue solid line). b) Shows all entries to the interaction Hamiltonian $H_I$ above $10^{-6}$. The width $W_0$, where $\Gamma = \Gamma(W_0)$ is shown in red. We can see that this does not extend past the coupling band. Shown for the impurity case, $J_z = B_x^{(S)} = 0$, with $N = 12$, $J_I = 0.4$. }
    \label{fig:Gamma_W}
\end{figure}
We thus define the trace over an energy width $W$, $\Tr_W \{\cdots\}$, as the trace over all states $\{|\phi_\alpha \rangle \}$ satisfying $W \geq |E_\alpha - E_\beta|$. This gives us $\Gamma$ as a function of the energy width $W$
\begin{equation}\label{eq:Gamma_W}
\Gamma(W) =  \pi \frac{\Tr_W \{H_I^\dagger H_I\}}{N^{*2}}D(E),
\end{equation}
where $N^{*2}$ is the number of elements included in $\Tr_W \{\cdots\}$.

The question is, then, which is the physically relevant value, $\Gamma(W_0)$, of the possible values of $\Gamma(W)$?
We know that $\Gamma$ must satisfy $\Gamma(W_0) \ll W_{BW}$, as otherwise our assumption that the density of states does not change over the width $\Gamma$ is invalid. Furthermore we must have $\Gamma(W_0) \ll W_0$ such that all states within the coupling energy $\Gamma := \Gamma(W_0)$ are counted. Thus we have the condition $\Gamma(W_0) \ll W_0 \ll W_{BW}$. We should expect to see a plateau in the function $\Gamma(W)$, giving the width over which the interaction Hamiltonian is effectively described by a random matrix. As $W$ grows we should then expect to see $\Gamma(W)$ decay for $W > W_{BW}$, as the long range interaction terms vanish. It is the value of $\Gamma(W)$ on the plateau that is the physically relevant point, as assuming the interaction strength is weak enough, the structure of long-range interactions should not matter. 

We can see from Fig. (\ref{fig:Gamma_W}) that this description is a good approximation for the spin chain, however the estimation of $\Gamma$ from this method is a likely source of error for the system sizes available, as the plateau region is not exactly flat as one would expect from a true random matrix. For larger sizes, one expects the initial structure of the Hamiltonian to be more `washed out' by the change to the non-interacting basis. We can also see from Fig. (\ref{fig:Gamma_W}) that as the interaction strength $J_I$ increases the line $\Gamma(W) = W$ will extend further into the plateau region, as the average value of the interaction Hamiltonian elements in this region increases. The random matrix approximation becomes invalid in the limit where the line $\Gamma(W) = W$ extends past the plateau region, as it is in this case that the density of states begins to change significantly over the width $\Gamma$ (hence the condition $\Gamma(W_0) \ll W_0$).

\begin{figure}
    \includegraphics[width=0.95\textwidth]{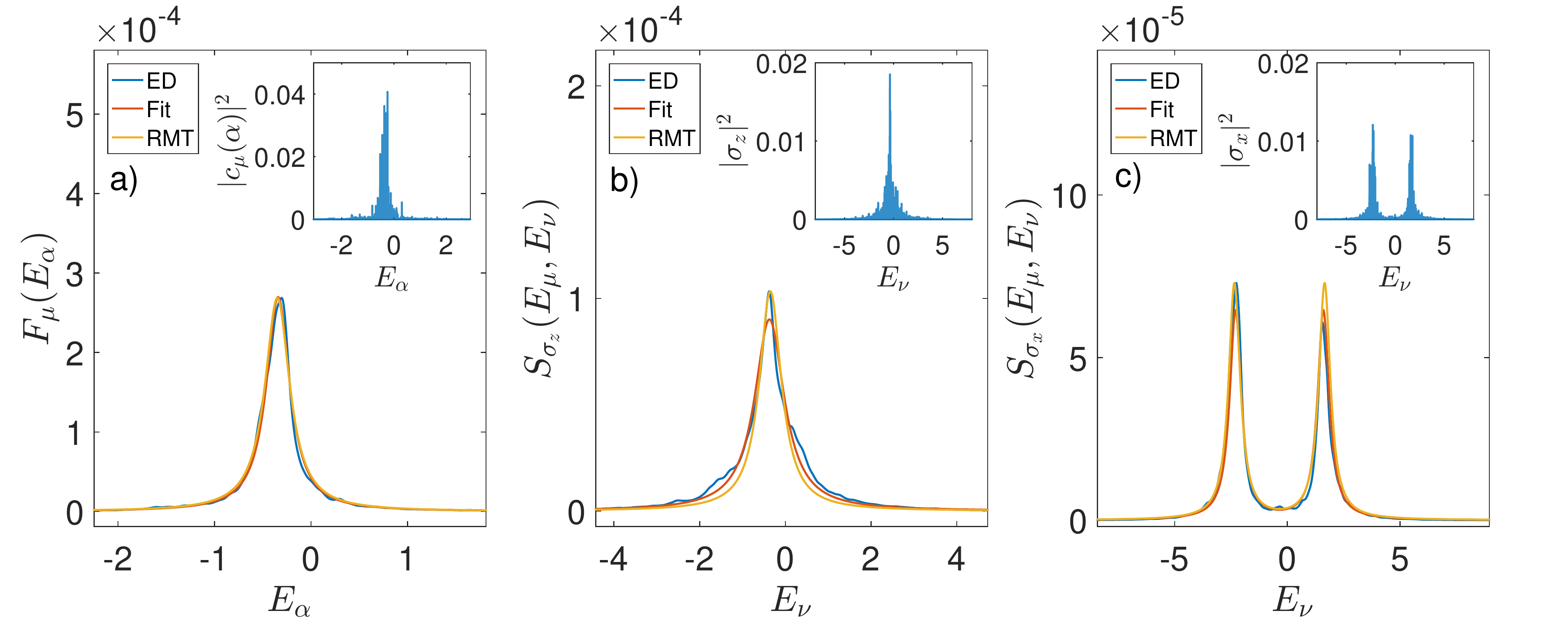}
    \caption{Smoothed ED calculation (blue) of the central eigenstate (a)), and off-diagonal elements of $|\sigma_z|^2$ (b)) and $|\sigma_x|^2$ (c)). Fit to a Eqs. (\ref{eq:Lambda}), (\ref{eq:O_sig_Z}), and (\ref{eq:O_sig_X}) respectively (red) and RMT prediction (yellow) using $\Gamma = \Gamma(W_0)$ (Eq. (\ref{eq:Gamma_W})) also shown for each. Raw data for each shown in insets. Each plot shown for $N = 13, N_S = 5, J_I = 0.5$, and for an energy $E_\mu$ in the centre of the spectrum.}
    \label{fig:ED_Examples_0}
\end{figure}
\subsection{Impurity}

\begin{figure}
    \includegraphics[width=0.99\textwidth]{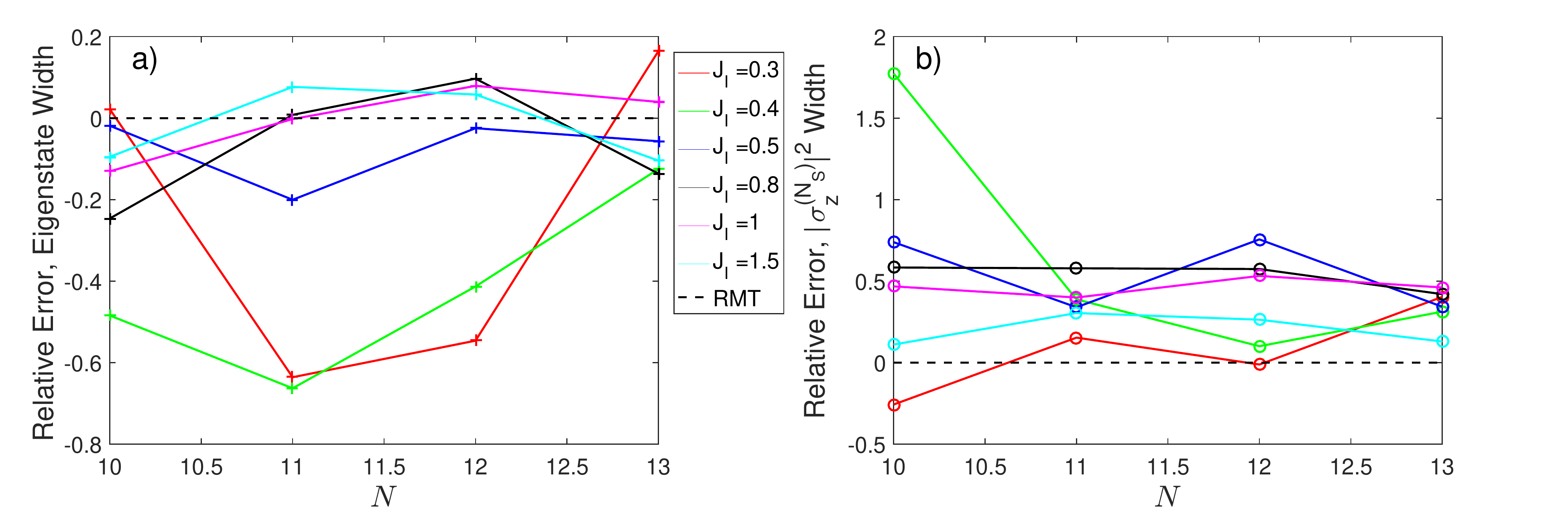}
    \caption{Comparison of fitted $\Gamma_{\text{Fit}}$ values with $\Gamma = \Gamma(W_0)$ (Eq. (\ref{eq:Gamma_W})). Relative error of each distribution width given by $(\Gamma_\text{Fit} - \Gamma(W_0)) / \Gamma(W_0)$. Comparisons shown for the central eigenstate ($\mu = 2^{N-1}$) for the fit to smoothed eigenstate distribution (a) and the $\sigma_z^{(N_S)}$ observable (b).}
    \label{fig:sigma_Z_err}
\end{figure}

We begin by analyzing the simple case where $J_z = B_x^{(S)} = 0$. Here the system qubit behaves differently to the bath, and thus can be thought of as an impurity. The natural observables in this case are the Pauli operators the $\sigma_z^{(N_S)}$ and $\sigma_x^{(N_S)}$.
It is straightforward to obtain the expected distribution for $\sigma_z^{(N_S)}$ from RMT by directly applying Eq. (\ref{eq:O_fin}), obtaining
\begin{equation}\label{eq:O_sig_Z}
|(\sigma_z^{(N_S)})_{\mu\nu}|^2_{\mu \neq \nu} = \frac{\omega_0 2\Gamma / \pi}{(E_\mu - E_\nu) + (2\Gamma)^2}.
\end{equation}
For the case of the $\sigma_x^{(N_S)}$ observable we must instead use information we have about the structure of the observable in the non-interacting basis to obtain a functional form for the observable distribution from the RMT formalism above. The useful observation here is that for $\sigma_x^{(N_S)}$
\begin{equation}
(\sigma_x^{(N_S)})_{\alpha\beta}  =\begin{cases}
    1, & \text{if $E_\beta = E_\alpha \pm 2B_z^{(S)}$}\\
    0, & \text{Otherwise}.
  \end{cases}
\end{equation}
Thus, using Eq. (\ref{eq:O_full}), and writing $\Lambda(\mu, \alpha) \to \Lambda(E_\mu, E_\alpha)$ for clarity, we find
\begin{equation}{\label{eq:O_sig_X}}
\begin{split}
|(\sigma_x^{(N_S)})_{\mu\nu}|^2_{\mu \neq \nu} & = \frac{1}{2}\int \frac{dE_\alpha}{\omega_0}\Lambda(E_\mu, E_\alpha)[\Lambda(E_\nu, E_\alpha + 2B_z^{(S)}) + \Lambda(E_\nu, E_\alpha - 2B_z^{(S)})] + \mathcal{O}(\omega_0^2) \\ & = \frac{\omega_0 \Gamma / \pi}{(E_\mu - E_\nu + 2B_z^{(S)}) + (2\Gamma)^2} + \frac{\omega_0 \Gamma / \pi}{(E_\mu - E_\nu - 2B_z^{(S)}) + (2\Gamma)^2},
\end{split}
\end{equation}
where the factor of $\frac{1}{2}$ is necessary for correct normalization. We can thus see that for the $\sigma_x^{(N_S)}$ observable we expect two peaks in the distribution of off-diagonal matrix elements, each of width $\Gamma$, separated by a width $4B_z^{(S)}$.

Shown in Fig. (\ref{fig:ED_Examples_0}) is a comparison between the ED numerical calculation and the RMT prediction. We compare the value for $\Gamma_{\text{Fit}}$ obtained from the fit to the smoothed distribution and the value found for $\Gamma(W_0)$ to obtain a relative error, shown in Fig. (\ref{fig:sigma_Z_err}), which we observe to decrease on average with system size for varying interaction strengths $J_I$. Whilst the range in relative error here is high, this is largely due to the difficulty in estimating $\Gamma$ for the available system sizes, and the fit to a Lorentzian distribution is very good. Furthermore, one would expect a high error for such short spin-chains, as the RMT result is valid in the thermodynamic limit, and requires the wave-function to be spread out over many states.
\subsection{Homogeneous Chain}
\begin{figure}
    \includegraphics[width=0.95\textwidth]{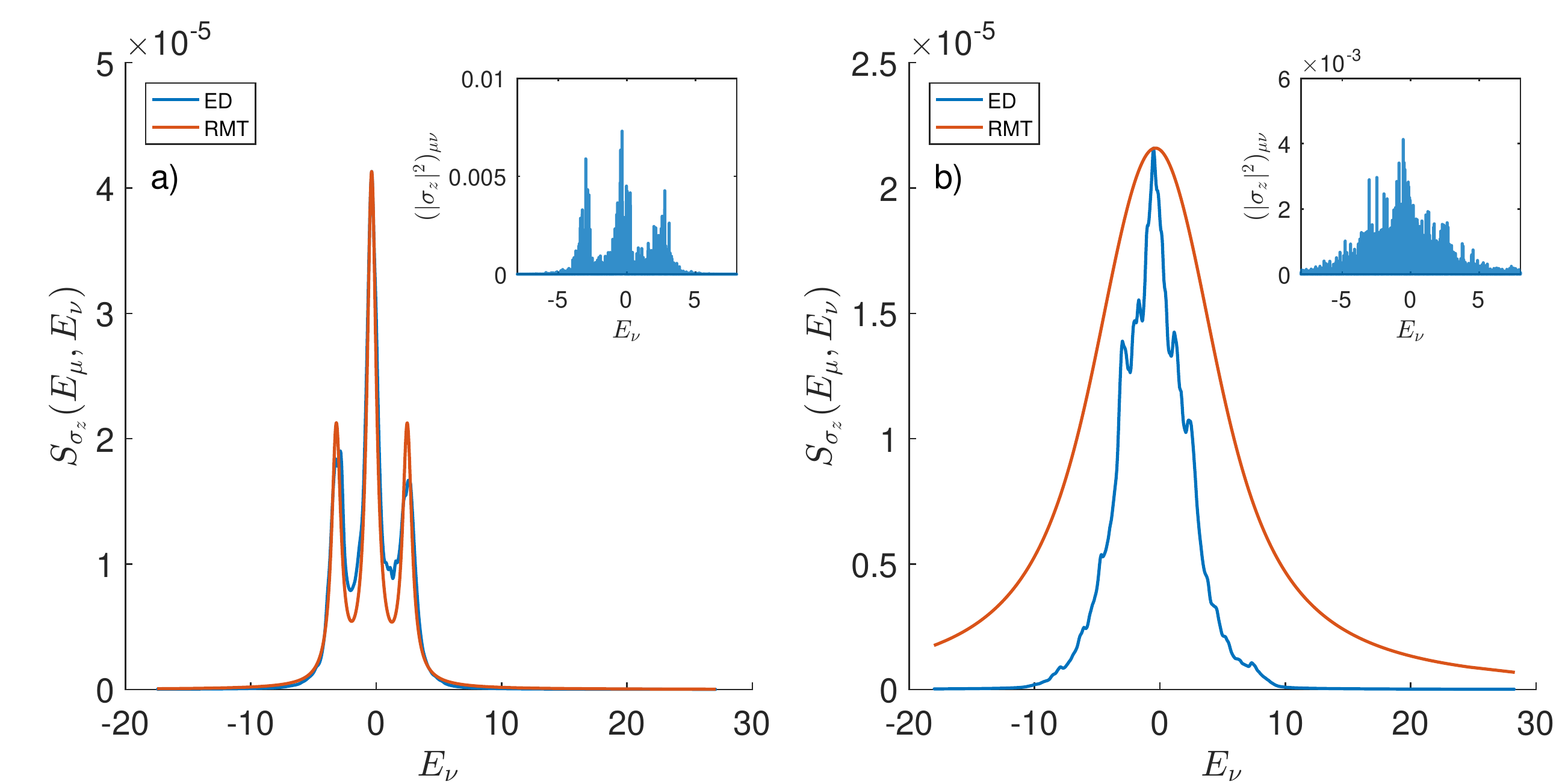}
    \caption{Smoothed ED calculation (blue) (raw data shown in insets) and RMT distribution (red) of smoothed $\sigma_z^{(N_S)}$ observable (analogous to Eq. (\ref{eq:O_sig_X}))  for couplings $J_I = J_z = 0.3$ (a) and the fully homogeneous case (b) $J_I = J_z = 1$. Each has $N = 13$,  $B_x^{(S)} = 1$, and $E_\mu$ in the centre of the spectrum.}
    \label{fig:sigma_Z_hom}
\end{figure}
The inclusion of a finite $B_x^{(S)}$ adds a level of complexity to the problem, as neither relevant observable  $\sigma_z^{(N_S)}$ or $\sigma_x^{(N_S)}$ is diagonal in the non-interacting basis. A similar approach to that shown in above for $\sigma_x^{(N_S)}$ allows us to calculate a distribution for $|\sigma_z^{(N_S)}|^2$, which for $B_x^{(S)} = 1$ one would expect to be made up instead of three Lorentzian peaks at $ E_\mu = E_\nu, E_\nu \pm 2\sqrt{2}$, with the central peak of twice the height.

\begin{figure}
    \includegraphics[width=0.95\textwidth]{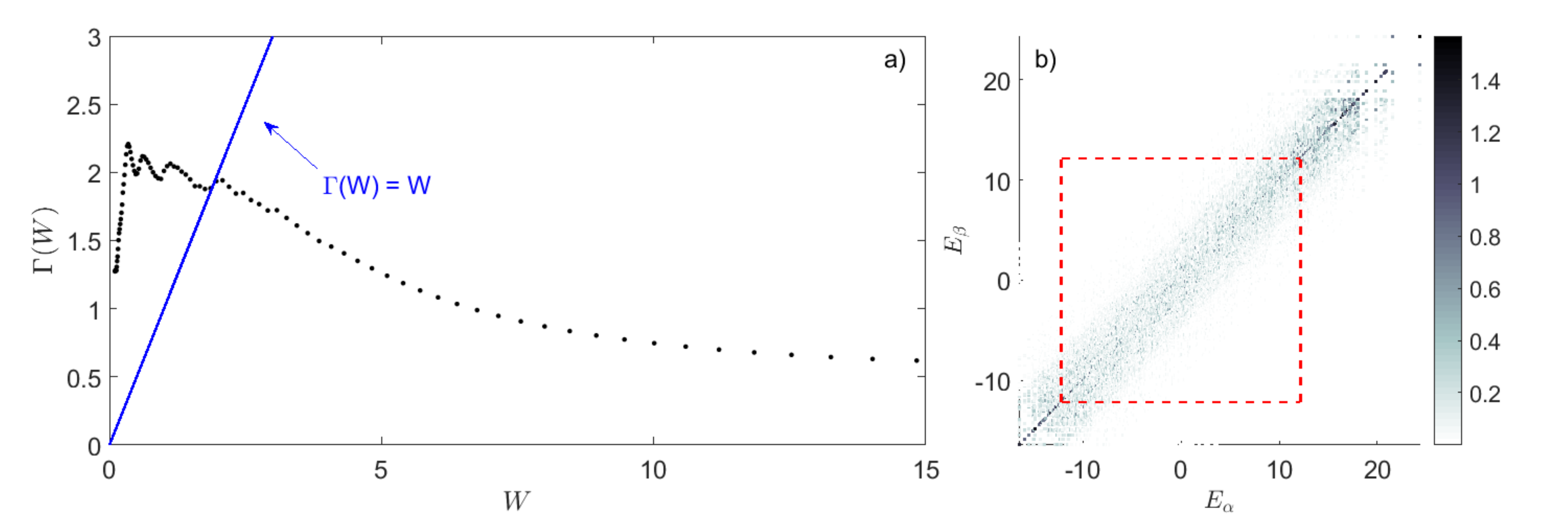}
    \caption{a) Plot of $\Gamma(W)$ (Eq. (\ref{eq:Gamma_W})) for varying values of $W$ (black dots), found using central half of the Hamiltonian energies only. We see that the $\Gamma(W) = W$ line (blue solid line) now extends much further, and thus the change in $\Gamma(W)$ value over this width alters the wave-function lineshape. b) Shows all entries to the interaction Hamiltonian $H_I$ above $10^{-6}$. Elements used in $\Gamma(W)$ calculation outlined by red dashed line, as elements outside this line retain structure due to finite size. Shown for $N = 12$, $B_x^{(S)} = 1$, $J_I = J_z = 1$. }
    \label{fig:Gamma_W_homogeneous}
\end{figure}

We can see in Fig. (\ref{fig:sigma_Z_hom}a) that we obtain a good agreement for the weak coupling case. As we approach the fully homogeneous case in Fig. (\ref{fig:sigma_Z_hom}b), however, we observe the RMT prediction no longer holds. We can see from Fig. (\ref{fig:Gamma_W_homogeneous}) that the $\Gamma(W) = W$ line extends to the end of the plateau region, and thus the requirements for assuming a full random matrix perturbation are not fulfilled - the change in density of states also contributes to the distribution of the wave-functions. Thus in this limit we no longer expect the wave-function distribution to be a Lorentzian, nor do we expect the method outlined above to be a good indication of the distribution width.

Furthermore we note that the for high couplings there are also added technical challenges for the systems available to our study, as the interaction Hamiltonian structure is not sufficiently randomized by the transformation to the non-interacting basis. We note that most of this structure occurs at the edges of the spectrum, and thus one can simply take the trace over the central half of the energies, as indicated in Fig. (\ref{fig:Gamma_W_homogeneous}b). This is justified for the bulk states we are analyzing.

\section{Finite Size Scaling of Long Time Fluctuations}
\begin{figure}
    \includegraphics[width=0.95\textwidth]{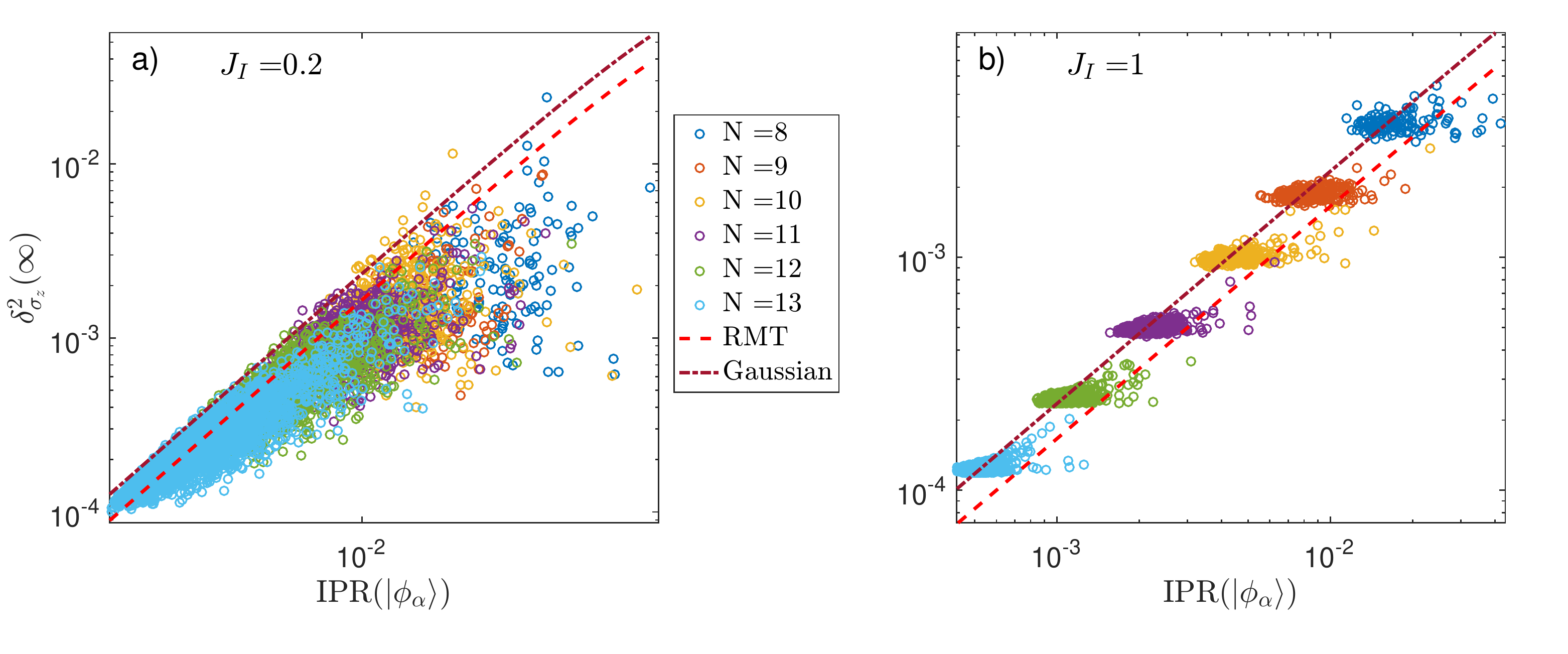}
    \caption{$\delta_{\sigma_z}^2(\infty)$ versus $\text{IPR}(|\phi_\alpha\rangle)$ for the central half of the spectrum of $\alpha$ values for the impurity case $B_x^{(S)} = J_z = 0$. a) Shows $J_I = 0.2$ b) Shows $J_I = 1$. Analytic result from RMT, Eq. (\ref{eq:Flucs2}) shown by red dashed line. Dash-dotted burgundy line shows prefactor obtained if $\Lambda(\mu, \alpha)$ is replaced by a Gaussian. }
    \label{fig:IPR_Fluc_ED}
\end{figure}
Off-diagonal elements of observables dominate the behaviour of their long-time fluctuations\cite{Srednicki1996, Srednicki1999, Beugeling2015}. Indeed, the infinite time (diagonal ensemble) fluctuations of an observable are given by
\begin{equation}\label{eq:Flucs}
\delta_O^2(\infty) = \sum_{\substack{\mu\nu \\ \mu\neq\nu}} |c_\mu(\alpha)|^2|c_\nu(\alpha)|^2|O_{\mu\nu}|^2 .
\end{equation}
Using the RMT result above, we may evaluate Eq. (\ref{eq:Flucs}) by the previous prescription of converting the sums to integrals, and integrating using the functional forms for $|c_\mu(\alpha)|^2$ and $|O_{\mu\nu}|^2_{\mu\neq\nu}$ derived above. We obtain 
\begin{equation}
\delta_O^2(\infty) = \frac{\omega_0}{4\pi\Gamma} \overline{[\Delta O_{\alpha\alpha}^2]}_{\overline{\mu}}  + \mathcal{O}(\omega_0^2)
\end{equation}
where the $O(\omega_0^2)$ term is due to the subtraction of the $\mu = \nu$ part. A further parameter that is of interest\cite{Clos2016a} to the finite size scaling of closed quantum systems is the Inverse Participation Ratio (IPR), defined as $\text{IPR}(\ket{\phi_\alpha}) = \sum_{\mu}|c_\mu(\alpha)|^4$. This can also be obtained in a similar manner using the RMT result
\begin{equation}
\text{IPR}(\ket{\phi_\alpha}) = \frac{3\omega_0}{2\pi\Gamma},
\end{equation}
where the factor of 3 in the denominator comes from the ratio of the second and fourth order moments of Gaussian variables. From these we obtain
\begin{equation}\label{eq:Flucs2}
\delta_O^2(\infty) = \frac{1}{6}\overline{[\Delta O_{\alpha\alpha}^2]}_{\overline{\mu}}\text{IPR}(\ket{\phi_\alpha}).
\end{equation}

\begin{figure}
    \includegraphics[width=0.95\textwidth]{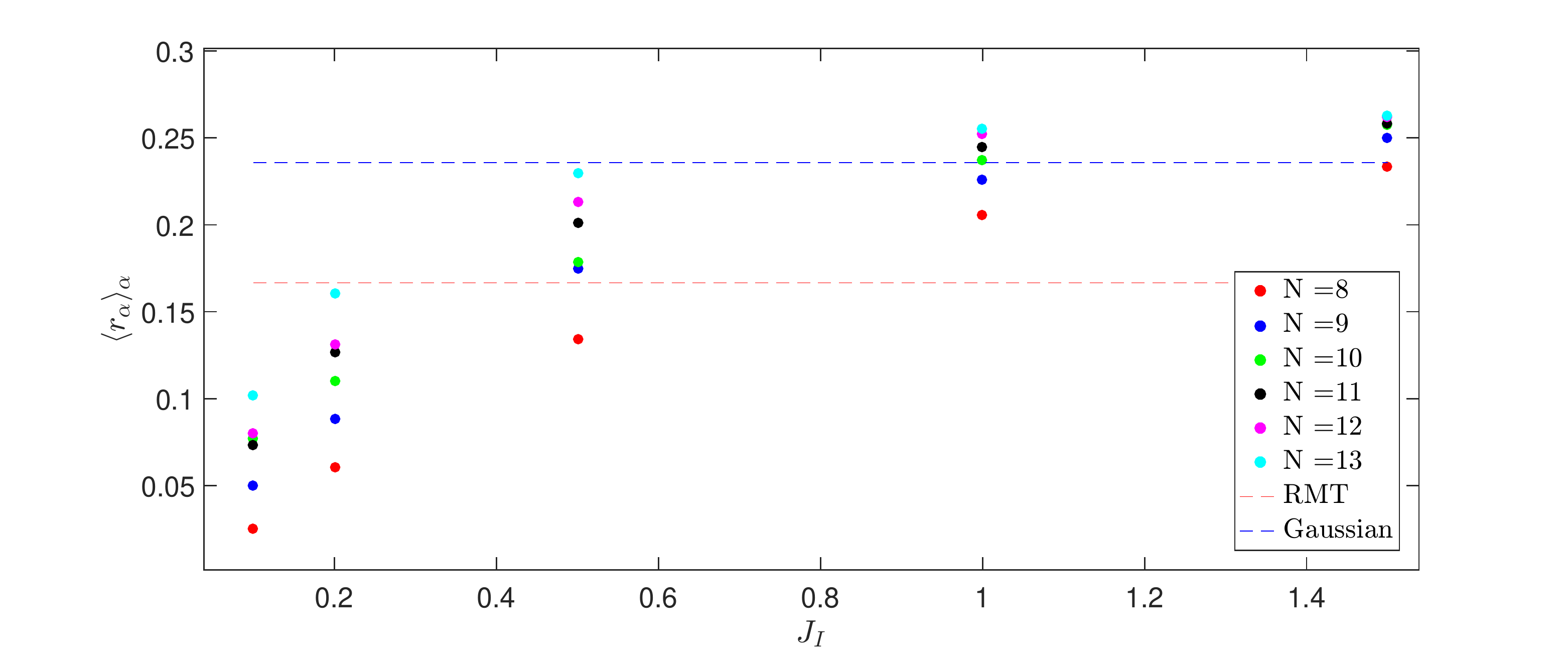}
    \caption{Plot of $\langle r_\alpha \rangle_\alpha$ (Eq. (\ref{eq:r_alpha})) as coupling $J_I$ is increased for impurity case $B_x^{(S)} = J_z = 0$. Average $\langle \cdots \rangle_\alpha $ taken over central 201 elements.}
    \label{fig:r_alpha_impurity}
\end{figure}

\begin{figure}
    \includegraphics[width=0.95\textwidth]{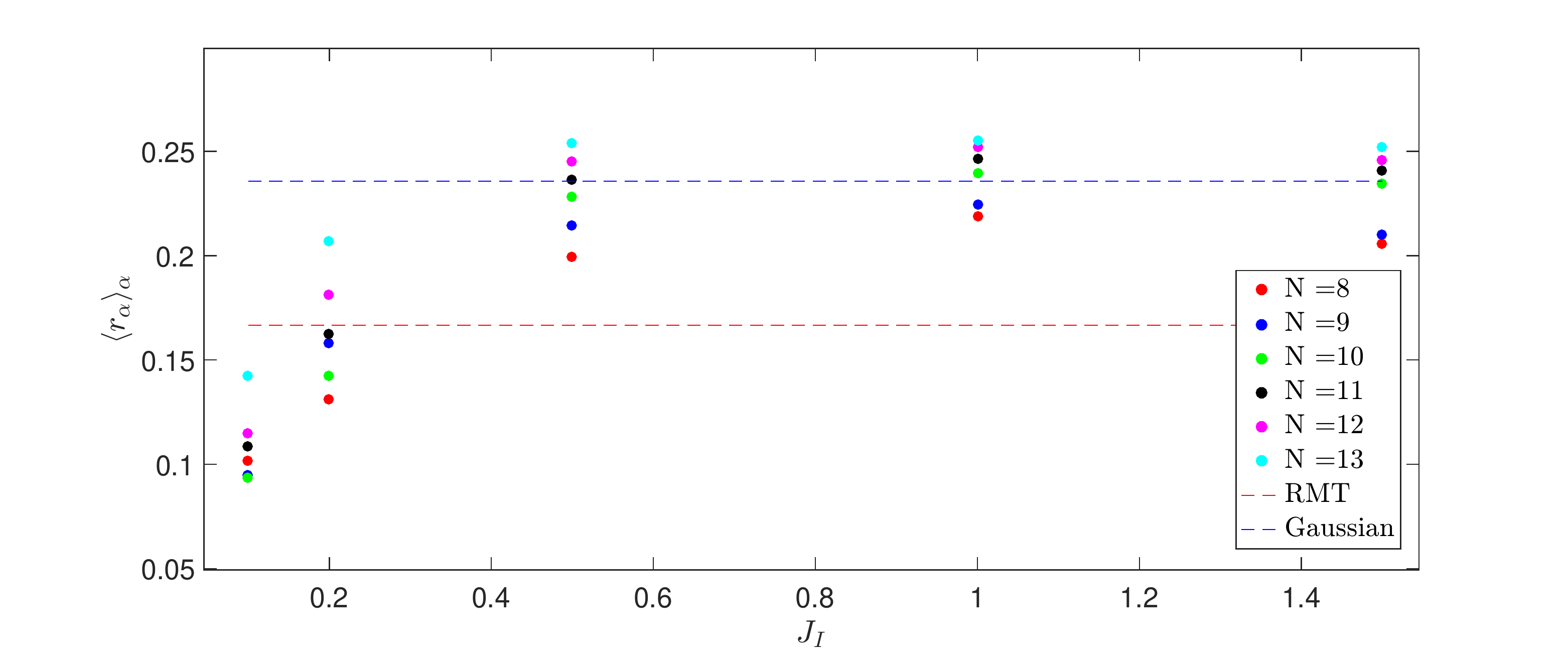}
    \caption{Plot of $\langle r_\alpha \rangle_\alpha$ (Eq. (\ref{eq:r_alpha})) as coupling $J_I$ is increased for homogeneous case $B_x^{(S)} = 1, J_z = J_I$. Average $\langle \cdots \rangle_\alpha $ taken over central 201 elements.}
    \label{fig:r_alpha_homogeneous}
\end{figure}

We can see from Fig. (\ref{fig:IPR_Fluc_ED}) that the proportionality is indeed correct, which has previously been shown to be a consequence of the ETH in reference \cite{Clos2016a}. Similar results have also been previously observed in references \cite{Short2011, Short2012, Reimann2008}, which obtain bounds on the late time fluctuations in terms if the IPR. Our work implies that in those systems which can be well described by a random matrix ansatz, the IPR determines not only an upper bound, but also the scale of the time-fluctuations. Similar dependencies have also been observed numerically in reference \cite{Neuenhahn2012}. One also observes in Fig. (\ref{fig:IPR_Fluc_ED}) that the numerical prefactor, expected to be $1/6$ for small couplings, as $\overline{[\Delta O_{\alpha\alpha}^2]}_{\overline{\mu}} = 1$ for the $\sigma_z^{(N_S)}$ observable here, seems to depend on the coupling strength. Motivated by previous numerical studies\cite{Mondaini2017, Santos2015, Santos2012, Njema1996, Flambaum1998, Atlas2017}, observing wave-functions of non-integrable systems to be Gaussian for large coupling strengths, one may repeat a similar calculation to that leading to Eq. (\ref{eq:Flucs2}), however with $\Lambda(\mu, \alpha)$ replaced by a Gaussian. We then obtain a prefactor of $(3\sqrt{2})^{-1}$. We define
\begin{equation}\label{eq:r_alpha}
r_\alpha = \overline{[\Delta O_{\alpha\alpha}^2]}_{\overline{\mu}}^{-1} \delta_O^2(\infty) \text{IPR}(\ket{\phi_\alpha})^{-1}
\end{equation}
in order to more closely analyze the dependence of the numerical prefactor as the coupling strengths are altered. For the case of Figs. (\ref{fig:r_alpha_impurity}) and (\ref{fig:r_alpha_homogeneous}) we show the change in $\langle r_\alpha \rangle_\alpha$, that is $r_\alpha$ averaged over many bulk $\alpha$ values, for the impurity and homogeneous cases respectively. Here we have $\overline{[\Delta O_{\alpha\alpha}^2]}_{\overline{\mu}} = 1$, and thus $\langle r_\alpha \rangle_\alpha$ gives the value of the prefactor directly. We indeed observe a growth of this prefactor to $\sim (3\sqrt{2})^{-1}$, the value expected by applying a Gaussian distributed wave-function to the RMT approach above in both the impurity and homogeneous cases. We note that for low couplings the fact that $\langle r_\alpha \rangle_\alpha$ does not tend exactly to the expected value from RMT is not surprising, as this is where we are most limited by the Hilbert space sizes available to our study, and thus there is a high associated error in this limit. Similar phenomena are observed for a numerical random matrix model, where the high coupling limit is obtained by the replacement $\Lambda(\mu, \alpha) = \Lambda = 1/N$, the scaling of fluctuations for this case is analyzed in Appendix \ref{App:RM_Fluc}.

\section{Discussion}
We have analytically studied a random matrix Hamiltonian, Eq. (\ref{eq:Hamiltonian}), made up of a linear ensemble of states with random interactions, and expanded on previous work \cite{Deutsch1991, Deutscha} to find a functional form for generic observables, as well as clarifying many of the approximations made to obtain the wave-function distribution (Appendix \ref{App:Lambda}). The form obtained for matrix elements of observables is in agreement with the ETH. We also predict that there is a linear relation between the time-fluctuations of an observable and the IPR. This relation may be relevant to detect quantum ergodicity by measuring the time-fluctuations in an experiment, if we understand quantum ergodicity as the participation of many Hamiltonian eigenstates in the initial state, which is implied by small IPR values.
Thus, measuring an exponential decrease of the time-fluctuations with system size would yield evidence that the IPR itself is exponentially decreasing with system size, which could be used as a smoking gun of quantum ergodicity.

We have assumed that an approximate description of the quantum dynamics of a subsystem in a many-body system can be achieved by an interaction term given by a structureless random Gaussian matrix. This approximation implies that the typical energy bandwidth of the coupling term, $W_{BW}$, is considered infinite compared to the coupling strength, $W_{BW} \gg \Gamma_\alpha$. Our results are thus immediately applicable to the stuation of an impurity weakly coupled to a many-body bath, since in this case $\Gamma_\alpha$ depends on a different interaction strength ($J_{I}$ in the spin chain example above) than the energy bandwidth, $W_{BW}$, and thus $\Gamma_\alpha$ can be made arbitrarily small. Our numerical calculations confirm that in this weak coupling limit many-body wave-functions are well approximated by  Lorentzian-shaped random wave-functions.

The weak coupling approximation may fail if, for example, we consider a subsystem in a homogeneous system where the coupling strength is not necessarily small compared to the bandwidth of the coupling term. Actually, in a homogeneous system we expect that $W_{BW} \approx \Gamma_\alpha$ since both energy scales are governed by the same interactions. For example, in the spin chain considered in the last section, both $W_{BW}$ and $\Gamma_\alpha$ are determined by the spin-spin interactions in the bulk $J_B$. 
In this case, we have observed numerically that the random wave-functions envelope is not necessarily a Lorentzian, but rather a Gaussian function. However, a valid random wave-function relying in the approximate distribution \eqref{eq:approx_prob_dist} is still possible by considering that 
$\Lambda(\mu,\alpha)$ are now normalized Gaussian functions. Most of the discussion in the subsequent sections remains intact, including expression \eqref{eq:offdiag} for the non-diagonal matrix elements of an observables. The only effect from the strong coupling condition is a different line-shape of envelope functions 
defined in Eqs. (\ref{eq:F}, \ref{eq:SO}), and a different prefactor in the scaling of the time-fluctuations as a function of the IPR, as shown in our numerical calculations (see Figs. (\ref{fig:r_alpha_impurity}) and (\ref{fig:r_alpha_homogeneous})). \\

We acknowledge funding by the People Programme (Marie Curie Actions) of the EU’s Seventh Framework Programme under REA Grant Agreement No. PCIG14-GA-2013-630955.


\appendix

\section{Variational Calculation of RMT Wavefunction Distribution}\label{App:Lambda}

To find the distribution of eigenstates for the random matrix system we must obtain a functional form of $\Lambda(\mu, \alpha)$ by minimising Eq. (\ref{eq:FreeEnergy}). 
Note that the integral in Eq. (\ref{eq:FreeEnergy}) is taken over all elements 
of $\{c_\mu(\alpha)\}$, \emph{i.e} 
$\int dc \to \prod_{\mu\alpha}\int dc_\mu(\alpha)$. 
The original probability distribution for the random wave-functions is given by Eq. \eqref{eq:int_dH}, which  may be re expressed by writing the second delta-function in Fourier form
\begin{equation}
\begin{split}
P(c)  = A\delta(cc^T - I)\int & \int \exp\bigg[-\sum_{\substack{\alpha^\prime\beta^\prime\\ \alpha^\prime>\beta^\prime}}\frac{N h_{\alpha^\prime\beta^\prime}^2}{2g_1^2} -\sum_{\alpha^\prime}\frac{N h_{\alpha^\prime\alpha^\prime}^2}{2g_2^2} - i \sum_{\substack{\mu^\prime\nu^\prime\\ \mu^\prime > \nu^\prime}}\lambda_{\mu^\prime\nu^\prime}\Big( \sum_{\substack{\alpha^\prime\beta^\prime\\\alpha^\prime > \beta^\prime}}c_{\mu^\prime}(\alpha^\prime)h_{\alpha^\prime\beta^\prime}c_{\nu^\prime}(\beta^\prime)  + \sum_{\substack{\alpha^\prime\beta^\prime\\\beta^\prime > \alpha^\prime}}c_{\mu^\prime}(\alpha^\prime)h_{\alpha^\prime\beta^\prime}c_{\nu^\prime}(\beta^\prime) \\ & + \sum_{\alpha^\prime} c_{\mu^\prime}(\alpha) h_{\alpha^\prime\alpha^\prime} c_{\nu^\prime}(\alpha) + \sum_{\alpha^\prime}c_{\mu^\prime}(\alpha^\prime)f_{\alpha^\prime}c_{\nu^\prime}(\alpha^\prime) \Big) \bigg] \bigg( \prod_{\substack{\alpha\beta \\ \alpha\geq \beta}}dh_{\alpha\beta}  \bigg)\bigg(  \prod_{\substack{\mu\nu\\ \mu>\nu}} d\lambda_{\mu\nu} \bigg),
\end{split}
\end{equation}
where we have expressed the independent widths of the off-diagonal and diagonal element distributions as $g_1$ and $g_2$ respectively. This further differs from that used in \cite{Deutscha} by appropriate symmetrization of the random interaction Hamiltonian. This may be rewritten as
\begin{equation}
\begin{split}
& P(c)  = A\delta(cc^T - I) \int   \int  \exp\bigg[-\sum_{\substack{\alpha^\prime\beta^\prime \\ \alpha^\prime\geq \beta^\prime}}\Bigg(\frac{N h_{\alpha^\prime\beta^\prime}^2}{2g_1^2}(1 - \delta_{\alpha^\prime\beta^\prime}) - \frac{N h_{\alpha^\prime\alpha^\prime}^2}{2g_2^2}\delta_{\alpha^\prime\beta^\prime} - i \sum_{\substack{\mu^\prime\nu^\prime\\ \mu^\prime > \nu^\prime}} \lambda_{\mu^\prime\nu^\prime}\bigg( \Big(c_{\mu^\prime}(\alpha^\prime)h_{\alpha^\prime\beta^\prime}c_{\nu^\prime}(\beta^\prime) + \\ & c_{\mu^\prime}(\beta^\prime)h_{\alpha^\prime\beta^\prime}c_{\nu^\prime}(\alpha^\prime)\Big)(1 - \delta_{\alpha^\prime\beta^\prime}) +  c_{\mu^\prime}(\alpha) h_{\alpha^\prime\alpha^\prime} c_{\nu^\prime}(\alpha)\delta_{\alpha^\prime \beta^\prime}\bigg)\Bigg) - i \sum_{\substack{\mu^\prime\nu^\prime\\ \mu^\prime > \nu^\prime}} \lambda_{\mu^\prime\nu^\prime}\sum_{\alpha^\prime}c_{\mu^\prime}(\alpha^\prime)f_{\alpha^\prime}c_{\nu^\prime}(\alpha^\prime)  \bigg] \bigg(\prod_{\substack{\alpha\beta \\ \alpha\geq \beta}} dh_{\alpha\beta} \bigg) \bigg(  \prod_{\substack{\mu\nu\\ \mu>\nu}} d\lambda_{\mu\nu} \bigg).
\end{split}
\end{equation}
The Gaussian integrals over $h_{\alpha\beta}$ may then be performed, giving
\begin{equation}
\begin{split}
P(c) = A^\prime \delta(cc^T - I) \int  \exp\bigg[ & - \sum_{\substack{\alpha^\prime\beta^\prime \\ \alpha^\prime\geq \beta^\prime}}\Bigg( \frac{g_1^2}{2N} \bigg(\sum_{\substack{\mu^\prime\nu^\prime\\ \mu^\prime > \nu^\prime}} \lambda_{\mu^\prime\nu^\prime} \Big(c_{\mu^\prime}(\alpha^\prime)c_{\nu^\prime}(\beta^\prime) + c_{\mu^\prime}(\beta^\prime)c_{\nu^\prime}(\alpha^\prime)\Big)\bigg)^2(1 - \delta_{\alpha^\prime \beta^\prime}) \\ & - \frac{g_2^2}{2N} \bigg(\sum_{\substack{\mu^\prime\nu^\prime\\ \mu^\prime > \nu^\prime}} \lambda_{\mu^\prime\nu^\prime} c_{\mu^\prime}(\alpha^\prime)c_{\nu^\prime}(\beta^\prime) \bigg)^2\delta_{\alpha^\prime \beta^\prime} \Bigg)- i\sum_{\substack{\mu^\prime\nu^\prime\\ \mu^\prime>\nu^\prime}} \lambda_{\mu^\prime\nu^\prime}\sum_{\alpha}c_{\mu^\prime}(\alpha)f_{\alpha}c_{\nu^\prime}(\alpha) \bigg]\prod_{\substack{\mu\nu\\ \mu>\nu}} d\lambda_{\mu\nu},
\end{split}
\end{equation}
where we have absorbed any constant prefactors into the new constant $A^\prime$. Now, the above equation may be transformed into a Gaussian integral by noting the following expansion of the first term in the exponent,
\begin{equation}\label{eq:expansion}
\begin{split}
 \sum_{\substack{\alpha^\prime\beta^\prime \\ \alpha^\prime \geq \beta^\prime}}&\sum_{\substack{\mu^\prime\nu^\prime\\ \mu^\prime > \nu^\prime}} \sum_{\substack{\mu\nu\\ \mu >\nu}} \lambda_{\mu\nu}\lambda_{\mu^\prime\nu^\prime} \Big(c_{\mu}(\alpha^\prime)c_{\nu}(\beta^\prime) + c_{\mu}(\beta^\prime)c_{\nu}(\alpha^\prime)\Big)\Big(c_{\mu^\prime}(\alpha^\prime)c_{\nu^\prime}(\beta^\prime) + c_{\mu^\prime}(\beta^\prime)c_{\nu^\prime}(\alpha^\prime)\Big)(1 - \delta_{\alpha^\prime \beta^\prime}) \\ & = \frac{1}{2}\sum_{\substack{\mu^\prime\nu^\prime\\ \mu^\prime > \nu^\prime}} \sum_{\substack{\mu\nu\\ \mu > \nu}} \lambda_{\mu\nu}\lambda_{\mu^\prime\nu^\prime} \sum_{\substack{\alpha^\prime\beta^\prime}}  \Big(c_{\mu}(\alpha^\prime)c_{\nu}(\beta^\prime)c_{\mu^\prime}(\alpha^\prime)c_{\nu^\prime}(\beta^\prime) + c_{\mu}(\beta^\prime)c_{\nu}(\alpha^\prime)c_{\mu^\prime}(\alpha^\prime)c_{\nu^\prime}(\beta^\prime) + \\ & \quad \quad c_{\mu}(\alpha^\prime)c_{\nu}(\beta^\prime)c_{\mu^\prime}(\beta^\prime)c_{\nu^\prime}(\alpha^\prime) + c_{\mu}(\beta^\prime)c_{\nu}(\alpha^\prime)c_{\mu^\prime}(\beta^\prime)c_{\nu^\prime}(\alpha^\prime) \Big) - 2 \sum_{\substack{\mu^\prime\nu^\prime\\ \mu^\prime > \nu^\prime}} \sum_{\substack{\mu\nu\\ \mu > \nu}} \lambda_{\mu\nu}\lambda_{\mu^\prime\nu^\prime} \sum_{\alpha^\prime} c_{\mu}(\alpha^\prime)c_{\nu}(\alpha^\prime)c_{\mu^\prime}(\alpha^\prime)c_{\nu^\prime}(\alpha^\prime) \\ & = \sum_{\substack{\mu^\prime\nu^\prime\\ \mu^\prime > \nu^\prime}} \sum_{\substack{\mu\nu\\ \mu > \nu}} \lambda_{\mu\nu}\lambda_{\mu^\prime\nu^\prime} \Big( \delta_{\mu \mu^\prime}\delta_{\nu\nu^\prime} + \delta_{\nu\mu^\prime}\delta_{\mu\nu^\prime}  \Big)- 2 \sum_{\substack{\mu^\prime\nu^\prime\\ \mu^\prime > \nu^\prime}} \sum_{\substack{\mu\nu\\ \mu > \nu}} \lambda_{\mu\nu}\lambda_{\mu^\prime\nu^\prime} \sum_{\alpha^\prime} c_{\mu}(\alpha^\prime)c_{\nu}(\alpha^\prime)c_{\mu^\prime}(\alpha^\prime)c_{\nu^\prime}(\alpha^\prime)  \\ & = \sum_{\substack{\mu\nu\\ \mu > \nu}} \lambda_{\mu\nu}^2 - 2 \sum_{\substack{\mu^\prime\nu^\prime\\ \mu^\prime > \nu^\prime}} \sum_{\substack{\mu\nu\\ \mu > \nu}} \lambda_{\mu\nu}\lambda_{\mu^\prime\nu^\prime} \sum_{\alpha^\prime} c_{\mu}(\alpha^\prime)c_{\nu}(\alpha^\prime)c_{\mu^\prime}(\alpha^\prime)c_{\nu^\prime}(\alpha^\prime) .
\end{split}
\end{equation}
Where in the last step we have used that $\lambda_{\mu\nu} = \lambda_{\nu\mu}$. Now, we have, assuming $g_2 = \sqrt{2}g_1 = \sqrt{2}g$, such that the random matrix perturbation is selected from the GOE,
\begin{equation}
\begin{split}
P(c) &= A^\prime \delta(cc^T - I)  \int  \exp\bigg[ - \frac{g_1^2}{2N} \sum_{\substack{\mu^\prime\nu^\prime\\ \mu^\prime > \nu^\prime}} \lambda_{\mu^\prime\nu^\prime}^2  + 2 \frac{g_1^2}{2N}\sum_{\substack{\mu^\prime\nu^\prime\\ \mu^\prime > \nu^\prime}} \sum_{\substack{\mu\nu\\ \mu > \nu}} \lambda_{\mu\nu}\lambda_{\mu^\prime\nu^\prime} \sum_{\alpha^\prime} c_{\mu}(\alpha^\prime)c_{\nu}(\alpha^\prime)c_{\mu^\prime}(\alpha^\prime)c_{\nu^\prime}(\alpha^\prime)\\ & \quad \quad - \frac{g_2^2}{2N} \sum_{\substack{\mu^\prime\nu^\prime\\ \mu^\prime > \nu^\prime}} \sum_{\substack{\mu\nu\\ \mu > \nu}} \lambda_{\mu\nu}\lambda_{\mu^\prime\nu^\prime} \sum_{\alpha^\prime} c_{\mu}(\alpha^\prime)c_{\nu}(\alpha^\prime)c_{\mu^\prime}(\alpha^\prime)c_{\nu^\prime}(\alpha^\prime)- i\sum_{\substack{\mu^\prime\nu^\prime\\ \mu^\prime>\nu^\prime}}\lambda_{\mu^\prime\nu^\prime}\sum_{\alpha}c_{\mu^\prime}(\alpha)f_{\alpha}c_{\nu^\prime}(\alpha) \bigg]\prod_{\substack{\mu\nu\\ \mu>\nu}} d\lambda_{\mu\nu} \\ &
= A^\prime \delta(cc^T - I) \prod_{\substack{\mu\nu\\ \mu>\nu}} \int  \exp\bigg[ - \frac{g^2}{2N}  \lambda_{\mu\nu}^2  - i\lambda_{\mu\nu}\sum_{\alpha}c_{\mu}(\alpha)f_{\alpha}c_{\nu}(\alpha) \bigg] d\lambda_{\mu\nu}.
\end{split}
\end{equation}
Carrying out the second Gaussian integral over $\lambda_{\mu\nu}$ we have,
\begin{equation}
P(c) = A^{\prime\prime}\delta(cc^T - I)  \exp{\Bigg[- \sum_{\substack{\mu\nu\\ \mu>\nu}}\frac{N \big(\sum_{\alpha}c_{\mu}(\alpha)f_{\alpha}c_{\nu}(\alpha)\big)^2}{2g^2} \Bigg]}.
\end{equation}
We note here that this leaves us with the same integral as would be obtained if we had enforced orthogonality of only two eigenvectors at a time, as in reference \cite{Deutscha}, up to a factor of two. Now, we observe

\begin{equation}\label{eq:P1}
\begin{split}
-\sum_{\substack{\mu\nu\\ \mu>\nu}} (\sum_{\alpha}c_{\mu}(\alpha)f_{\alpha}c_{\nu}(\alpha))^2 & = -\frac{1}{2}\sum_{\substack{\mu\nu\\ \mu\neq\nu}} (\sum_{\alpha}c_{\mu}(\alpha)f_{\alpha}c_{\nu}(\alpha))^2\\& = -\frac{1}{2}\sum_{\mu\nu}(\sum_{\alpha}c_{\mu}(\alpha)f_{\alpha}c_{\nu}(\alpha))^2 + \frac{1}{2}\sum_{\mu}(\sum_{\alpha}f_{\alpha}c_{\mu}^2(\alpha))^2 \\& =  - \frac{1}{2}\sum_{\alpha}f_\alpha^2 + \frac{1}{2}\sum_{\mu}(\sum_{\alpha}f_{\alpha}c_{\mu}^2(\alpha))^2,
\end{split}
\end{equation}
as $\sum_{\alpha}c_\mu(\alpha)c_\nu(\alpha) = \delta_{\mu\nu}$. We then have,
\begin{equation}
P(c) = A^{\prime\prime}\delta(cc^T - I)  \exp{\Bigg[- \frac{N\sum_{\alpha}f_{\alpha}^2}{4g^2} + \sum_{\mu}\frac{N \big(\sum_{\alpha}f_{\alpha}c_{\mu}^2(\alpha)\big)^2}{4g^2} \Bigg]}.
\end{equation}
Thus, we finally obtain
\begin{equation}
P(c) =  \frac{1}{Z_P}\Bigg(\prod_{\mu\nu} \delta(\sum_\alpha c_\mu(\alpha)c_\nu(\alpha)) \Bigg)\Bigg(\prod_{\mu} \delta(\sum_\alpha c_\mu^2(\alpha) - 1)\Bigg) \exp{\bigg[\frac{N}{4g^2}\sum_{\mu}(\sum_\alpha f_\alpha c_{\mu}^2(\alpha))^2 \bigg]},
\end{equation}
where we have absorbed the constant terms into a new constant $1/Z_P$, and written explicitly the full form of the delta-function $\delta(cc^T - I)$.
One can see by Gibbs' inequality, $p(c, \Lambda) \ln{\frac{P(c)}{p(c, \Lambda)}} \leq 0$, that we can obtain the best possible approximation $p(c, \Lambda)$ by obtaining the functional form of $\Lambda$ that fulfils
\begin{equation}
\frac{\partial F}{\partial\Lambda} = 0,
\end{equation}
as well as any constraints on $\Lambda$ we may require. This is the problem solved in reference \cite{Deutscha}, though using a different target distribution $p(c, \Lambda)$. The Free energy integral of Eq. (\ref{eq:FreeEnergy}) can be split into two parts, which we heuristically label the `energy', $E$, and `entropy', $S$, with $F = E - S$, we have
\begin{equation}{\label{eq:Energy1}}
E = - \int p(c, \Lambda) \ln{\frac{\Big(\prod_{\mu} \delta(\sum_\alpha c_\mu^2(\alpha) - 1) \Big)\exp{\bigg[\frac{N}{4g^2}\sum_{\mu}(\sum_\alpha f_\alpha c_{\mu}^2(\alpha))^2\bigg]}}{Z_P}} dc
 \end{equation}
and
\begin{equation}{\label{eq:Entropy1}}
S = - \int p(c, \Lambda) \ln{\frac{\exp{\bigg[-\sum_{\mu\alpha}\frac{c_\mu^2(\alpha)}{2\Lambda(\mu, \alpha)}\bigg]}}{Z_p}} dc.
\end{equation}
Note that the orthogonality condition delta-functions in $P(c)$ and $p(c, \Lambda)$ cancel in Eq. (\ref{eq:FreeEnergy}) to obtain the above expressions for $S$ and $E$. To calculate the Free Energy we need to evaluate the partition function $Z_p$, which is given by
\begin{equation}
Z_p = \int\int   \exp{\bigg[-\sum_{\mu\alpha}\frac{c_{\mu}^2(\alpha)}{2\Lambda(\mu, \alpha)} - i\sum_{\substack{\mu\nu\\ \mu > \nu}}\lambda_{\mu\nu}\sum_{\alpha}c_{\mu}(\alpha)c_{\nu}(\alpha)\bigg]}\left( \prod_{\mu, \alpha}dc_\mu(\alpha)\right) \left(\prod_{\substack{\mu\nu \\ \mu > \nu}}d \lambda_{\mu\nu}\right),
\end{equation}
where we have written the delta-function as a Fourier integral. The condition $\mu > \nu$ is required such that pairwise interactions are not doubly counted. From the second equality one may recognise that this integral may be seen as an average over a Gaussian distribution of $c_\mu(\alpha)$s, thus we write 
\begin{equation}{\label{eq:Z_p1}}
Z_p = \int  Z_G\bigg\langle \exp{\bigg[ - i\sum_{\substack{\mu\nu\\ \mu > \nu}}\lambda_{\mu\nu} \sum_{\alpha}c_{\mu}(\alpha)c_{\nu}(\alpha) \bigg]} \bigg\rangle_G  \prod_{\substack{\mu\nu\\\mu >\nu}}d\lambda_{\mu\nu},
\end{equation}
where we have defined
\begin{equation}
\langle A \rangle_G := \frac{1}{Z_G}  \int A \exp{\bigg[-\sum_{\mu\alpha}\frac{c_\mu^2(\alpha)}{2\Lambda(\mu, \alpha)}\bigg]}\prod_{\mu\alpha}dc_\mu(\alpha),
\end{equation}
with
\begin{equation}
Z_G :=\int \exp{\bigg[-\sum_{\mu\alpha}\frac{c_\mu^2(\alpha)}{2\Lambda(\mu,\alpha)}\bigg]} \prod_{\mu\alpha} dc_\mu(\alpha) = \prod_{\mu\alpha}(2\pi\Lambda(\mu, \alpha))^{\frac{1}{2}}.
\end{equation}
Expanding the exponent, we may write
\begin{equation}\label{eq:Z_p2}
Z_p = \int  Z_G \Bigg\langle \sum_{n=0}^\infty \frac{(-i)^n}{n!}\bigg(\sum_{\substack{\mu\nu\\\mu >\nu}}\lambda_{\mu\nu}O_{\mu\nu}\bigg)^n \Bigg\rangle_G\prod_{\substack{\mu\nu\\\mu >\nu}}d\lambda_{\mu\nu} ,
\end{equation}
where $O_{\mu\nu} := \sum_{\alpha} c_\mu(\alpha)c_\nu(\alpha)$. From this we can see each term in the power series of Eq. (\ref{eq:Z_p2}) scales as the average of increasing powers of the operator $O$, summed over the eigenstates. As odd moments of $O$ are identically zero we are immediately left with only even terms. Furthermore, as the average is taken with a Gaussian distribution of $c_\mu(\alpha)$s, the only non-zero terms in the average occur when labels are equal in pairs. For example, for the first non-zero term ($n = 2$) we have
\begin{equation}
\langle O_{\mu\nu}O_{\mu^\prime\nu^\prime} \rangle_G = \langle \sum_{\alpha\alpha^\prime}c_{\mu}(\alpha)c_{\nu}(\alpha)c_{\mu^\prime}(\alpha^\prime)c_{\nu^\prime}(\alpha^\prime) \rangle_G \approx \langle \sum_{\alpha}c_{\mu}(\alpha)^2c_{\nu}(\alpha)^2 \rangle_G,
\end{equation}
so
\begin{equation}
\begin{split}
\sum_{\substack{\mu^\prime\nu^\prime\\ \mu^\prime > \nu^\prime}} \sum_{\substack{\mu\nu\\ \mu > \nu}} \lambda_{\mu \nu}\lambda_{\mu\prime\nu\prime} \langle O_{\mu\nu}O_{\mu\prime\nu\prime} \rangle_G &= \sum_{\substack{\mu^\prime\nu^\prime\\ \mu^\prime > \nu^\prime}} \sum_{\substack{\mu\nu\\ \mu > \nu}} \lambda_{\mu \nu}\lambda_{\mu^\prime\nu^\prime} \langle \sum_{\alpha\alpha^\prime}c_{\mu}(\alpha)c_{\nu}(\alpha)c_{\mu^\prime}(\alpha^\prime)c_{\nu^\prime}(\alpha^\prime) \rangle_G \\ & = \sum_{\substack{\mu\nu\\ \mu > \nu}} \lambda_{\mu \nu}^2 \langle \sum_{\alpha}c_{\mu}(\alpha)^2c_{\nu}(\alpha)^2 \rangle_G.
\end{split}
\end{equation}
We then obtain 
\begin{equation}
\langle \sum_{\alpha}c_{\mu}^2(\alpha)c_{\nu}^2(\alpha) \rangle_G = \frac{1}{Z_G} \int  \sum_{\alpha}c_{\mu}^2(\alpha)c_{\nu}^2(\alpha) \exp{\bigg[-\sum_{\mu\alpha}\frac{c_\mu^2(\alpha)}{2\Lambda(\mu, \alpha)}\bigg]}\prod_{\mu\alpha}dc_\mu(\alpha)  = \sum_\alpha \Lambda(\mu, \alpha) \Lambda(\nu, \alpha).
\end{equation}
%
%
To calculate the full average in Eq. (\ref{eq:Z_p2}), we need to calculate the average of all the even powers,
\begin{equation}
\langle (\sum_{\substack{\mu \nu \\ \mu > \nu}} O_{\mu,\nu} )^{2n} \rangle_G =  
\sum_{\substack{\mu_1 \nu_1 \\ \mu_1 > \nu_1}} 
\sum_{\substack{\mu_2 \nu_2 \\ \mu_2 > \nu_2}} \dots
\sum_{\substack{\mu_{2 n} \nu_{2 n} \\ \mu_{2 n} > \nu{2 n}}}  
\langle 
O_{\substack{\mu_1 \nu_1}} O_{\substack{\mu_2 \nu_2}} \dots O_{\substack{\mu_{2n} \nu_{2n}}}			
		\rangle .
\end{equation}
In the limit of large participation ratios, $\Gamma/\omega_0 \gg 1$, the dominant contribution comes from contractions between pairs, $O_{\mu_i, \nu_i}$ and $O_{\mu_j,\nu_j}$, such that,  
\begin{equation}
\sum_{\substack{\mu_1 \nu_1 \\ \mu_1 > \nu_1}} 
\sum_{\substack{\mu_2 \nu_2 \\ \mu_2 > \nu_2}} \dots
\sum_{\substack{\mu_{2 n} \nu_{2 n} \\ \mu_{2 n} > \nu_{2 n}}}  
\langle 
O_{\substack{\mu_1 \nu_1}} O_{\substack{\mu_2 \nu_2}} \dots O_{\substack{\mu_{2n} \nu_{2n}}}			
\rangle \approx 
(2n - 1) !! \Big(\sum_{\substack{\mu\nu\\\mu >\nu}}\lambda_{\mu\nu}^2\sum_{\alpha}\Lambda(\mu, \alpha)\Lambda(\nu, \alpha)\Big)^{n},
\end{equation}
where the factor $(2n - 1)!! := 1 \cdot 3 \cdot 5 \cdots (2n - 1)$, arises after the counting of all possible combinations of pairs. In this approximation we are neglecting those terms where indices are not contracted by pairs, however, those terms have at least one less summation over one of the indices $\alpha_j$, and thus they are supressed by a factor $(\Gamma/\omega_0)^{-1}$. 
We may thus re-express the average in (\ref{eq:Z_p1}) as 
\begin{equation}
\bigg\langle \sum_{n=0}^\infty \frac{(-i)^n}{n!}(\sum_{\substack{\mu\nu\\\mu >\nu}}\lambda_{\mu\nu}O_{\mu\nu})^n \bigg\rangle_G \approx \sum_{n=0}^\infty \frac{(-i)^{2n}}{(2n)!} (2n - 1)!! \Big(\sum_{\substack{\mu\nu\\\mu >\nu}}\lambda_{\mu\nu}^2\sum_{\alpha}\Lambda(\mu, \alpha)\Lambda(\nu, \alpha)\Big)^{n}.
\end{equation}
Now, as $(2n - 1)!! = \frac{(2n)!}{2^nn!}$ and $(-i)^{2n} = (-1)^n$, we can finally write
\begin{equation}{\label{eq:PartFuncFullApp}}
\begin{split}
Z_p & =  Z_G  \int \exp{\left[ -\frac{1}{2} \sum_{\substack{\mu\nu\\ \mu > \nu}} \lambda_{\mu\nu}^2\sum_{\alpha} \Lambda(\mu, \alpha)\Lambda(\nu, \alpha)  \right]}\prod_{\substack{\mu\nu\\\mu >\nu}}d\lambda_{\mu\nu} \\ & = (2\pi)^{N^2-N/2}\left(\prod_{\mu\alpha} (\Lambda({\mu},\alpha))^{\frac{1}{2}}\right) \left( \prod_{\substack{\mu\nu\\\mu >\nu}} \bigg(\sum_{\alpha} \Lambda(\mu, \alpha)\Lambda(\nu, \alpha) \bigg)^{-{\frac{1}{2}}}\right).
\end{split}
\end{equation}
Now, the entropy, Eq. (\ref{eq:Entropy1}) may be easily evaluated to obtain (noting once more that the integral is understood by $\int dc \to \prod_{\mu\alpha}\int dc_\mu(\alpha)$)
\begin{equation}
S = \sum_{\mu\alpha}\Lambda(\mu, \alpha)\frac{\partial Z_p}{\partial \Lambda(\mu, \alpha)} + \ln{Z_p}.
\end{equation}
Using the partition function found in Eq. (\ref{eq:PartFuncFullApp}), and ignoring any constant terms, which do not contribute to the final form of $\Lambda(\mu, \alpha)$, we write,
\begin{equation}
\ln (Z_p) = \frac{1}{2} \sum_{\mu\alpha} \ln(\Lambda(\mu, \alpha)) - \frac{1}{4} \sum_{\mu\nu} \ln (\sum_{\alpha}\Lambda(\mu,\alpha)\Lambda(\nu,\alpha))
\end{equation}
so
\begin{equation}
S = - \sum_{\mu\alpha}\frac{\Lambda(\mu,\alpha)}{2}\sum_{\mu^\prime\neq\mu}\frac{\Lambda(\mu^\prime,\alpha)}{\sum_{\alpha^\prime}\Lambda(\mu,\alpha^\prime)\Lambda(\mu^\prime,\alpha^\prime)} + \frac{1}{2} \sum_{\mu\alpha}\ln(\Lambda(\mu, \alpha)) - \frac{1}{4}  \sum_{\mu\nu} \ln (\sum_{\alpha}\Lambda(\mu,\alpha)\Lambda(\nu,\alpha)).
\end{equation}
Now, we can see that the first term is also constant, as the sum over $\alpha$ cancels that over $\alpha^\prime$, thus this term does not contribute, and we obtain
\begin{equation}
S = \frac{1}{2}\sum_{\mu\alpha}\ln{\Lambda(\mu, \alpha)} - \frac{1}{4}\sum_{\mu\nu} \ln{\sum_{\alpha}\Lambda(\mu, \alpha)\Lambda(\nu, \alpha)}.
\end{equation}
We note that in ref \cite{Deutscha} the first term is here labelled entropy, and the second term is labelled a repulsion energy. The final calculation required for evaluation of the free energy is the energy part, Eq. (\ref{eq:Energy1}). As $Z_P$ does not depend on $\Lambda(\mu,\alpha)$, we can ignore this part (as we require $F$ only for it's derivative with respect to $\Lambda(\mu, \alpha)$). We can also re-write the delta-function factor as $\prod_{\mu}\delta(\sum_{\alpha}c_\mu^2(\alpha) - 1) \to \lim_{\mathcal{U} \to \infty} \exp{[\mathcal{U}\sum_{\mu}(\sum_{\alpha}c_\mu^2(\alpha) - 1)^2]}$, and thus we are left with
\begin{equation}
\begin{split}
E = -\frac{1}{Z_p}  & \int \int  \exp{\bigg[-\sum_{\mu\alpha}\frac{c_{\mu}^2(\alpha)}{2\Lambda(\mu, \alpha)} - i\sum_{\substack{\mu \nu \\ \mu > \nu}} \lambda_{\mu\nu}\sum_{\alpha}c_{\mu}(\alpha)c_{\nu}(\alpha)\bigg]} \\ & \times \bigg[ \lim_{\mathcal{U} \to \infty} \mathcal{U}\sum_{\mu}(\sum_{\alpha}c_{\mu}^2(\alpha) - 1)^2 + \frac{N}{4g^2}\sum_{\mu}(\sum_{\alpha}f_\alpha c_\mu^2(\alpha))^2 \bigg]\left(\prod_{\mu\alpha}dc_\mu(\alpha)\right)\left( \prod_{\substack{\mu\nu \\ \mu > \nu}} d\lambda_{\mu\nu}\right)
\end{split}
\end{equation}
We can see that the first term here ensures normalization of the $c_\mu(\alpha)$s, and is zero provided this condition is met. The second term, similarly to the partition function, may be re-expressed as the Gaussian average
\begin{equation}{\label{eq:Energy2}}
E = -\frac{1}{Z_p} \int  Z_G \bigg\langle \frac{N}{4g^2}\sum_{\mu^\prime}(\sum_{\alpha^\prime}f_{\alpha^\prime} c_{\mu^\prime}^2(\alpha^\prime))^2  \sum_{n=0}^\infty \frac{(-i)^n}{n!}(\sum_{\substack{\mu\nu \\ \mu> \nu}}\lambda_{\mu\nu}O_{\mu\nu})^n \bigg\rangle_G\prod_{\substack{\mu\nu\\ \mu > \nu }}d\lambda_{\mu\nu}.
\end{equation}
Now, as with the partition function this cannot be calculated exactly, but we can use the fact that the average is  taken over a Gaussian distribution of $c_\mu(\alpha)$s to find the dominant part. This is most clearly seen by writing the average in the form
\begin{equation}\label{eq:Gau_ave_E}
\bigg\langle \frac{N}{4g^2}\sum_{\mu^\prime}(\sum_{\alpha^\prime\alpha^{\prime\prime}}f_{\alpha^\prime} f_{\alpha^{\prime\prime}}   c_{\mu^\prime}^2(\alpha^\prime) c_{\mu^{\prime}}^2(\alpha^{\prime\prime}))  \sum_{n=0}^\infty \frac{(-i)^n}{n!}(\sum_{\substack{\mu\nu \\ \mu> \nu}}\lambda_{\mu\nu}\sum_{\alpha}c_\mu(\alpha) c_\nu(\alpha))^n \bigg\rangle_G.
\end{equation}
The key observation here is that as the only non-zero terms in Eq. (\ref{eq:Gau_ave_E}) are those with even powers of $c_\mu(\alpha)$ any terms that have correlations between the `bias' factor $H_b = \frac{N}{4g^2} \sum_{\mu}(\sum_{\alpha}f_\alpha c_\mu^2(\alpha))^2$ from the Hamiltonian and the orthogonality factor are either excluded by the fact that $\mu \neq \nu$ or reduced by the need for $\alpha = \alpha^\prime, \alpha^{\prime\prime}$. Thus the dominant cause of correlations, leading to non-zero terms in the average, are from correlations \emph{within} each factor, and not between. This leads to the approximation, which is equivalent to that made in the partition function evaluation above,
\begin{equation}
\begin{split}
E & \approx \frac{1}{Z_p}  \int  Z_G \bigg\langle\frac{N}{4g^2} \sum_{\mu^\prime}(\sum_{\alpha^\prime}f_{\alpha^\prime} c_{\mu^\prime}^2(\alpha^\prime))^2 \exp{\bigg[-\frac{1}{2}\sum_{\substack{\mu\nu \\ \mu> \nu}}\lambda_{\mu\nu}^2\sum_{\alpha} \Lambda(\mu, \alpha) \Lambda(\nu, \alpha) \bigg]}\bigg\rangle_G\prod_{\substack{\mu\nu \\ \mu> \nu}}d\lambda_{\mu\nu} \\ & = -\bigg\langle \frac{N}{4g^2} \sum_{\mu}(\sum_{\alpha}f_{\alpha} c_{\mu}^2(\alpha))^2 \bigg\rangle_G \\ & := -\langle H_b \rangle_G.
\end{split}
\end{equation}
Explicitly, we have
\begin{equation}
\begin{split}
E = & - \frac{1}{Z_G}  \int  \frac{N}{4g^2}\sum_{\mu}(\sum_{\alpha}f_{\alpha} c_{\mu}^2(\alpha))^2 \exp{\bigg[-\sum_{\mu\alpha}\frac{c_\mu^2(\alpha)}{2\Lambda(\mu, \alpha)}\bigg]}\prod_{\mu\alpha}dc_\mu(\alpha).
\end{split}
\end{equation}
We see that the contributions of the averages $\langle c_\mu^2(\alpha)\rangle_G$ and $\langle c_\mu^4(\alpha)\rangle_G$ may be ignored, as they are proportional to $\Lambda(\mu, \alpha)^{3/2}$ and $\Lambda(\mu, \alpha)^{5/2}$ respectively, and are thus small. We may therefore approximate this as
\begin{equation}
\begin{split}
E \approx & - \frac{1}{Z_G} \frac{N}{4g^2}\sum_{\mu}(\sum_{\alpha}f_{\alpha} c_{\mu}^2(\alpha))^2 \int   \exp{\bigg[-\sum_{\mu\alpha}\frac{c_\mu^2(\alpha)}{2\Lambda(\mu, \alpha)}\bigg]}\prod_{\mu\alpha}dc_\mu(\alpha)  \\ & = - \frac{N}{4g^2}\sum_{\mu}(\sum_{\alpha}f_{\alpha} c_{\mu}^2(\alpha))^2 \\ & \approx -\frac{N}{4g^2}\sum_{\mu}(\sum_{\alpha}f_{\alpha} \Lambda(\mu, \alpha))^2,
\end{split}
\end{equation}
where the last step is valid provided that the function $\Lambda(\mu, \alpha)$ is sufficiently smooth. We are now able to write the full functional form of the Free Energy from $F = E - S$,
\begin{equation}\label{eq:FreeEnergyFull}
F = -\frac{N}{4g^2}\sum_{\mu}(\sum_{\alpha}f_{\alpha} \Lambda(\mu, \alpha))^2 + \frac{1}{4}\sum_{\mu\nu} \ln{\sum_{\alpha}\Lambda(\mu, \alpha)\Lambda(\nu, \alpha)} - \frac{1}{2}\sum_{\mu\alpha}\ln{\Lambda(\mu, \alpha)}.
\end{equation}
Now, we wish to find the function $\Lambda(\mu, \alpha)$ that minimises Eq. (\ref{eq:FreeEnergyFull}) under the conditions $\sum_{\alpha}\Lambda(\mu, \alpha) = \sum_{\mu}\Lambda(\mu,\alpha) = 1$. We thus introduce the corresponding Lagrange multipliers into Eq. (\ref{eq:FreeEnergyFull}), and find the derivative with respect to $\Lambda(\mu^\prime, \alpha^\prime)$. We thus wish to find the functional form for $\Lambda(\mu, \alpha)$ satisfying
\begin{equation}
\frac{\partial F}{\partial \Lambda(\mu^\prime, \alpha^\prime)} + \eta_{\mu^\prime} + \eta_{\alpha^\prime} = 0,
\end{equation}
where we have introduced the Lagrange coefficients $\eta_{\mu(\alpha)}$ of their respective multipliers $\eta_{\mu(\alpha)}(\sum_{\mu(\alpha)}\Lambda(\mu, \alpha) - 1)$. Now, we have
\begin{equation}
\frac{\partial F}{\partial \Lambda(\mu^\prime, \alpha^\prime)} = -\frac{N}{2g^2}f_\alpha^\prime \sum_{\alpha} f_\alpha \Lambda(\mu^\prime, \alpha) + \frac{1}{2} \sum_{\nu} \frac{\Lambda(\nu, \alpha^\prime)}{\sum_{\alpha}\Lambda(\mu^\prime, \alpha)\Lambda(\nu, \alpha)} - \frac{1}{2}\frac{1}{\Lambda(\mu^\prime, \alpha^\prime)},
\end{equation}
which we may simplify given that due to the normalization condition we have `incompressibility'\cite{Deutscha} of bulk eigenstates, and thus 
\begin{equation}
\Lambda(\mu, \alpha) = \Lambda(\mu - \alpha),
\end{equation} 
and
\begin{equation}
\sum_{\alpha} \alpha \Lambda(\mu, \alpha) = \sum_{\alpha} \alpha \Lambda(\mu - \alpha) = \mu,
\end{equation} 
after a suitable change of variables. Thus we have 
\begin{equation}
\frac{\partial F}{\partial \Lambda(\mu^\prime, \alpha^\prime)} = -\frac{N\omega_0^2\alpha^\prime \mu^\prime}{2g^2} + \frac{1}{2} \sum_{\nu} \frac{\Lambda(\nu, \alpha^\prime)}{\sum_{\alpha}\Lambda(\mu^\prime, \alpha)\Lambda(\nu, \alpha)} - \frac{1}{2}\frac{1}{\Lambda(\mu^\prime, \alpha^\prime)}.
\end{equation}
Now, we make the ansatz
\begin{equation}
\Lambda(\mu, \alpha) = \frac{\omega_0 \Gamma / \pi}{(E_\mu - E_\alpha)^2 + \Gamma^2}.
\end{equation}
Taking the continuum limit and noting that 
\begin{equation}
\int \frac{dE_\alpha}{\omega_0}\Lambda(\mu, \alpha)\Lambda(\nu, \alpha) = \frac{2\omega_0 \Gamma / \pi}{(E_\mu - E_\nu)^2 + 4\Gamma^2}
\end{equation}
we thus obtain
\begin{equation}\label{eq:Lambda_condition}
-\frac{N\omega_0^2\alpha^\prime \mu^\prime}{2g^2} + \frac{\omega_0\pi}{4\Gamma}\mu^{\prime 2} + \frac{\omega_0\pi}{2\Gamma}\mu^\prime \alpha^\prime + \frac{\omega_0\pi}{2\Gamma}(\mu^{\prime 2} + \alpha^{\prime 2}) + \eta_{\mu^\prime} + \eta_{\alpha^\prime} = 0,
\end{equation}
where we have absorbed all constant terms into the Lagrange multipliers $\eta_{\mu^\prime}$ and $\eta_{\alpha^\prime}$. We now note that terms in $\alpha^\prime$ may be absorbed into the Lagrange multiplier $\eta_{\mu^\prime}$ and vice versa, thus we can readily observe that the condition Eq (\ref{eq:Lambda_condition}) is fulfilled for $\Gamma = \pi g^2 / N \omega_0$. 
\section{Scaling of Fluctuations for Random Matrix Hamiltonian}\label{App:RM_Fluc}
\begin{figure}
    \includegraphics[width=0.95\textwidth]{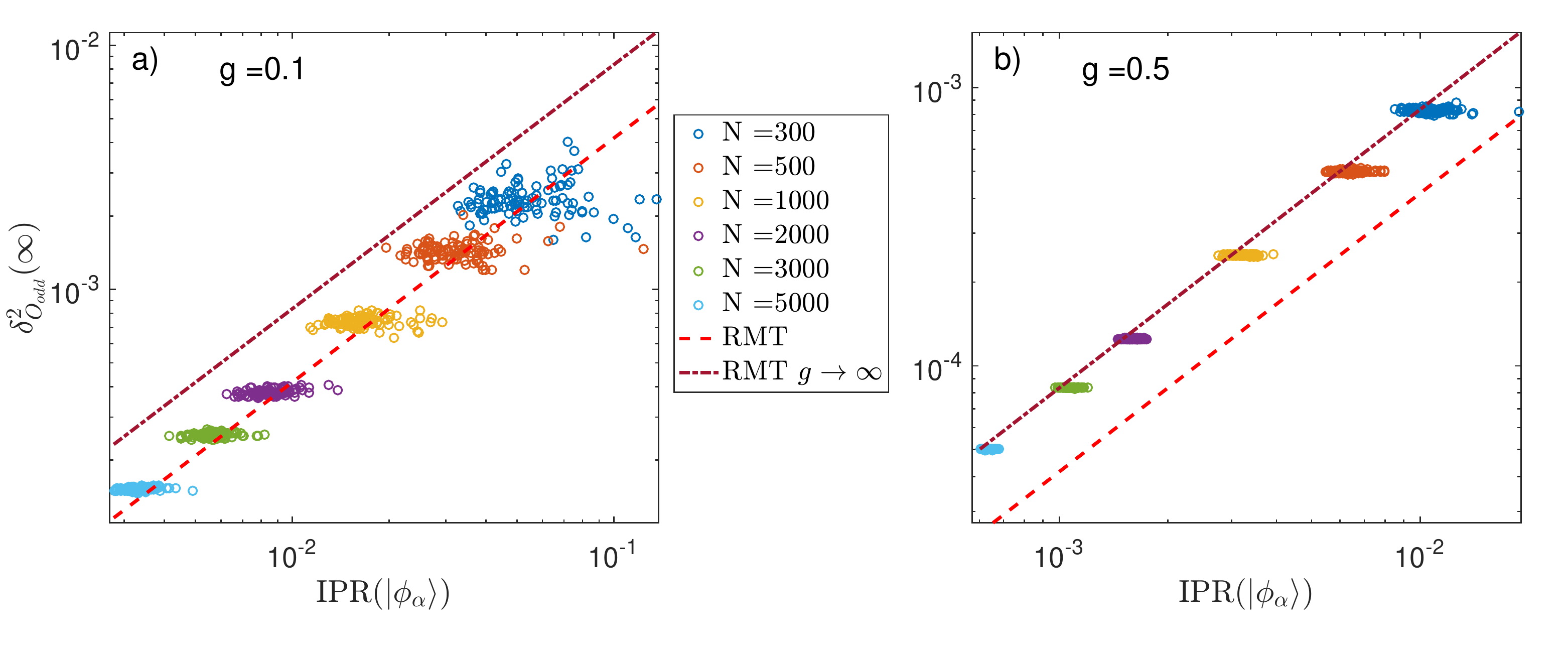}
    \caption{$\delta_{O_{\rm odd}}^2$ versus $\text{IPR}(|\phi_\alpha\rangle)$ for the central 100 values of $\alpha$ in the spectrum for a single realization (no averaging) of the random matrix system. a) Shows $g = 0.1$ b) Shows $g = 0.5$. Analytic result from RMT, Eq. (\ref{eq:Flucs2}) shown by red dashed line, $g \to \infty$ limit of RMT result, Eq. (\ref{eq:Flucs_inf}), shown by dash-dotted burgundy line.}
    \label{fig:IPR_Fluc_RM}
\end{figure}

If we analyze the $g \to \infty$ limit of our random matrix system, where we have simply a real Hermitian random matrix as our Hamiltonian. In this limit we thus expect to see $\Lambda(\mu, \alpha) = 1/N$. This can be easily seen to minimise our free energy, Eq. (\ref{eq:FreeEnergyFull}), in the $g \to \infty$ limit. Repeating, then, the analysis above, we find in this limit
\begin{equation}
\langle c_\mu(\alpha)c_\nu(\alpha)c_\mu(\beta)c_\nu(\beta) \rangle_V = \frac{1}{N^2}\delta_{\alpha\beta} - \frac{1}{N^3}(1 + \delta_{\alpha\beta}),
\end{equation}
from which we obtain
\begin{equation}
\begin{split}
|O_{\mu\nu}|^2_{\mu\neq\nu} & = \sum_{\alpha\beta} \bigg(\frac{1}{N^2}\delta_{\alpha\beta} - \frac{1}{N^3}(1 + \delta_{\alpha\beta})\bigg)O_{\alpha\alpha}O_{\beta\beta} \\
 & \approx \overline{[O_{\alpha\alpha}^2]}_{\overline{\mu}} \sum_\alpha \bigg(\frac{1}{N^2} - \frac{1}{N^3}\bigg) -  \overline{[O_{\alpha\alpha}]}_{\overline{\mu}}^2 \sum_{\alpha\beta}\frac{1}{N^3} \\
 & \approx \frac{ \overline{[\Delta O_{\alpha\alpha}^2]}_{\overline{\mu}}}{N}.
\end{split}
\end{equation}
Now, the infinite time fluctuations may now be obtained via Eq. (\ref{eq:Flucs}), from which we find
\begin{equation}
\delta_O^2(\infty) = \sum_{\mu\nu}\frac{1}{N^2}\frac{ \overline{[\Delta O_{\alpha\alpha}^2]}_{\overline{\mu}}}{N} \approx \frac{ \overline{[\Delta O_{\alpha\alpha}^2]}_{\overline{\mu}}}{N}.
\end{equation}
The IPR may be easily seen to be equal to $3 / N$, where once again the factor of three originates in the relationship between the second and fourth moments of Gaussian distributed variables. From this we obtain
\begin{equation}\label{eq:Flucs_inf}
\delta_O^2(\infty) = \frac{1}{3} \overline{[\Delta O_{\alpha\alpha}^2]}_{\overline{\mu}}\text{IPR}(\ket{\phi_\alpha}),
\end{equation}
\begin{figure}
    \includegraphics[width=0.95\textwidth]{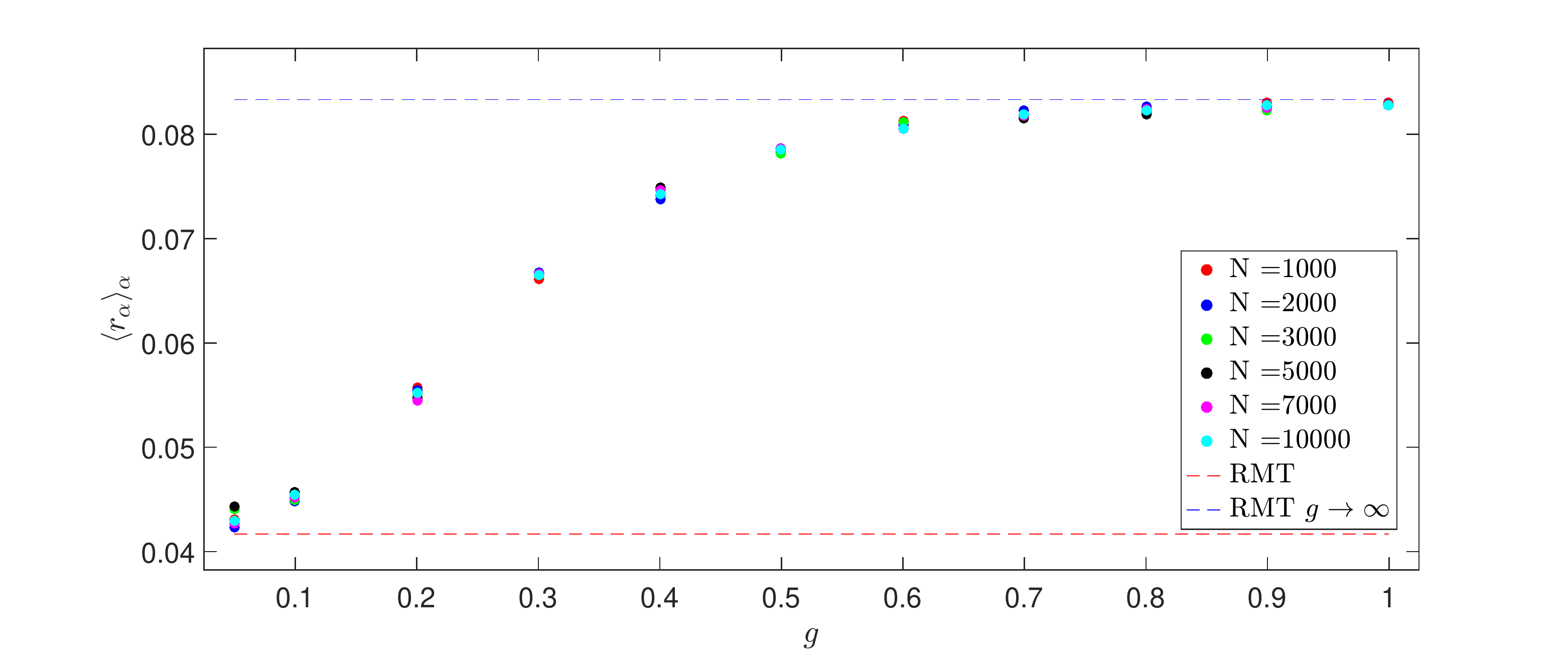}
    \caption{Plot of $\langle r_\alpha \rangle_\alpha$ (see Eq. (\ref{eq:r_alpha})) as coupling $g$ is increased for random matrix Hamiltonian of Eq. (\ref{eq:Hamiltonian}). Average $\langle \cdots \rangle_\alpha $ taken over central 201 elements.}
    \label{fig:r_alpha_RM}
\end{figure}
as the expected scaling of infinite time fluctuations for the $g \to \infty$ limit. Thus we can see from Fig. (\ref{fig:r_alpha_RM}) that the factor of two emerges from our RMT model when the coupling is large. 

\bibliographystyle{apsrev4-1}
\bibliography{bibli}

\end{document}